\documentstyle[aps,prb,preprint,epsf]{revtex}
\begin{document}

\title{Weighted two particle Green's functions in the CPA}
\draft
\author{N.F. Schwabe and R.J. Elliott}
\address{Department of Physics, University of Oxford \\
Theoretical Physics, 1 Keble Road, Oxford OX1 3NP, UK}
\date{\today}
\maketitle

\begin{abstract}
We extend the two particle theory of disordered systems
within the coherent potential approximation CPA to obtain weighted
contributions to averaged two particle
resolvents which arise from separate alloy components. Starting from
 first principles in a model of diagonal
disorder and the single site approximation for a binary substitutional
alloy $A_cB_{1-c}$ we extend the approach of a fundamental paper by
 Velick{\'y} to
evaluate various weighted forms of a
general class of two particle Green's functions.
Applications in a wide
 range of linear response
theory are discussed in detail as well as the behavior of the
weighted functions in a strong disorder limit. To exemplify our
 analytic
calculations the optical absorption in a disordered model-alloy
 is studied
numerically.
\end{abstract}
\vspace{3ex}

\section{Introduction}
In recent years the understanding of the effects of
disorder in the physics of metals, semiconductors and amorphous
 systems
has made a tremendous progress. This vigorous development was
motivated to a great extent by a thorough understanding of how
strongly
 disorder effects determine the behavior of real physical systems.
The success of the various analytical descriptions which have been
considered,
 however, has always been determined by the relative simplicity to
which
 approximations could be reduced, in order to keep the theory
tractable
 analytically, without losing its capability to account for the most
 important physical aspects.
One of the most successful approximations to match these requirements
has been the coherent potential approximation CPA developed by Soven
 \cite{bi1},
Taylor \cite{bi2}, and
extended by Leath \cite{bi7,bi8,bi9}, Velick{\'y}
\cite{bi3,bi4,bi5,bi6} and many others (see Ref. \onlinecite{bi13}
 for a review).

In contrast to many other approximations for the procedure of
 configurational averaging in disordered systems, the CPA is capable
 of interpolating correctly between the limits of
weak \cite{bi10} and strong disorder \cite{bi12} as well as low and
high impurity concentrations. Therefore it is also able to predict
 accurately
the formation of impurity bound states turning into split-off
impurity bands as the impurity concentration is increased.
Extensions of the single site CPA to include scattering from clusters
of impurities are still today the only existing analytic theories
 that allow for
a calculation of the density of states in disordered systems such that
a reasonable versimilitude is attained \cite{bi14}.
The main difficulty with the CPA, is the relative
complexity of the self consistent equations which have to be solved in
more accurate extensions of the theory.

In the present paper we propose a more differentiated  analysis of two
particle properties within the CPA building on the single site
approximation (SSA) in which it was developed originally. Our novel
treatment should be especially useful for applications to binary
substitutional alloys in  all regimes of disorder.
Following an earlier paper by Velick{\'y} \cite{bi4}, in which the
regular two particle theory within the CPA was first introduced and
which will be referred to hereafter as {\bf I},  the theory
of weighted single particle resolvents \cite{bi26} is extended to a
 properly
weighted two particle theory.
Weighting of a Green's function in this context means that through the
application of appropriate operators to the unaveraged resolvent,
restrictions are made on the type of alloy component of either
or both of the sites on which the
particle starts and terminates its motion. Upon
 averaging, this results in a statistical weight being attributed
 to the
averaged unrestricted resolvent with possibly separate additive
 terms.

On the average, therefore, contributions to a full two particle
 Green's
function in terms of constituent components can be resolved. This
allows for a better understanding of how several physical processes
contribute to a cumulative behavior as the principal parameters of the
system are varied and it can therefore be
used in an analysis of further effects which differentiates between
these components.

We have structured the paper as follows: In section II the
important features of the single particle CPA for diagonal disorder
are recapitulated and relations which are important to the calculation
of the corresponding two particle Green's functions are established.
 In
section III we calculate the Fourier transforms of a class of weighted
two particle functions which are kept as general as
possible. For that reason only a representative choice of weightings
 are
calculated explicitly, since other weights can be obtained in an
 analogous fashion.
Section IV is devoted to possible applications of weighted two
particle functions
in linear response theory, and the peculiarities of two different
classes of such functions are discussed in detail, which we have
selected to cover a large range of conceivable applications.

In the third part of section IV the behavior of one
class of functions discussed before is examined in a
split band limit (strong disorder). Section V is devoted to a
numerical
 study of the splitting into
several components of an interband absorption spectrum of a
disordered alloy. Section VI in conclusion discusses the implications
and possible further applications of the results obtained throughout
this work.

\section{Single particle properties}

We consider a binary substitutional alloy $A_c B_{1-c}$ on a simple
totally disordered monoatomic lattice on which each site is occupied
by either an $A-$ or a $B$-component with  probabilities $c$ and
 $1-c$, respectively.
Since by symmetry it is only necessary to consider concentrations
between $0 \leq c \leq 0.5$, we restrict our investigation to this
region and define $A$ as the impurity and $B$
as the host component of the alloy.

 To simplify  further calculations we
consider diagonal disorder only, i.e. a Hamiltonian of the form
\begin{equation}
H = H_0 + U = H_0 + \sum_n U_n \label{a0}
\end{equation}
where $H_0$ represents the periodic part and the
$U_n $ are single site contributions to a random disorder
potential which assume either the value $U^A_n = V$ or $U^B_n = 0$
with probabilities $c$ and $1-c$, depending
on whether the site $n$ is an impurity or part of the host material.
 In
principle one could also have introduced a symmetric model for the
 disorder,
but as it turns out the amount of algebra is somewhat reduced by the
asymmetric definition, while switching from one form to the other
 does not
present any difficulty.

The propagator of the disordered
medium $G$ relates to the one of a pure host medium (pure $B$-phase)
$g$ through
 the Dyson equation
\begin{equation}
G=g+gUG=g+gUg+gUGUg \label{a1}.
\end{equation}

The CPA for a disordered medium is introduced by the usual method
\cite{bi1,bi2} of placing the impurities in a  self-consistent
medium such that the the averaged propagator of this effective medium
fulfills the relation
\begin{equation}
\bar G=g+g \Sigma \bar G \label{a2}
\end{equation}
where $\Sigma$ is the CPA single particle self energy.
 $\Sigma$ itself is determined by the self
consistency condition that the average of the total scattering of a
particle in the effective medium be zero. This total scattering is
described by the equation
\begin{equation}
G=\bar G + \bar G T \bar G \label{a8}.
\end{equation}
where $T$ is corresponding T-matrix of the problem and the self
 consistency condition is therefore
$ \langle T \rangle \equiv 0$. In the SSA an additional requirement is
made through the condition that the total scattering off a single
site $n$ be zero. This scattering is described by the single site
contribution to the T-matrix $T_n$ which is defined as
\begin{equation}
T_n = (U_n - \Sigma_n) \left[ 1 + F T_n \right] \label{a3}
\end{equation}
where $F$ is the site diagonal average propagator and $U_n$ and
 $\Sigma_n$ are the single site decompositions of $U$
and $\Sigma$, such that $U= \sum_n U_n$ and $\Sigma_n = \sum_n
\Sigma_n$.
The average of the disorder potential $U$ alone amounts to $\langle U
\rangle = c V$ This along with the average of (\ref{a3}) set to zero
determines $\Sigma$ to be
\begin{equation}
\Sigma = \frac {cV} {1-(V- \Sigma )F} \label{a3a}.
\end{equation}

One can now define conditional or weighted propagators \cite{bi26}
 which
explicitly describe the propagation of a particle between partly or
 completely
specified types of sites by multiplying a normalized version of the
random potential onto them.
For example,
\begin{equation}
G^i = \frac {U G} V \label{a4}
\end{equation}
describes the motion of a particle commencing on an impurity site and
ending at
an arbitrary other site in the medium, since $U$ will be zero if the
first site of the function is a host. Similarly,
\begin{equation}
G^{ii}= \frac {UGU} {V^{2}} \label{a6}
\end{equation}
describes a situation where both sites are required to be
impurities for the function not to be zero.
Upon averaging over all configurations, the Green's functions
become translationally invariant and the following relationships
between the averaged weighted and unweighted functions can be obtained
\begin{eqnarray}
\bar G^i & = & \frac \Sigma V \bar G \label{a5} \\
\bar G^{ii} & = & \left( \frac \Sigma V \right)^2 \bar G + \frac
 {\Sigma - cV} {V^2}. \label{a7}
\end{eqnarray}
The second term corrects the site diagonal elements when $\bar G^{ii}
= \bar G^i$ and uses identity (\ref{a3a}). From here on, other
 weighted functions can be calculated by
probability conservation. It is found that
\begin{eqnarray}
\bar G^h & = & \bar G - \bar G^i = \left( 1- \frac \Sigma V \right)
  \bar
G \label{a5a} \\
\bar G^{ih} & = & \bar G^{hi} = \bar G^i - \bar G^{ii} =  \frac
 \Sigma V \left(
1 - \frac \Sigma V \right) \bar G - \frac {\Sigma - cV} {V^2}
\label{a7a} \\
\bar G^{hh} & = & \bar G^h - \bar G^{ih} = \left( \frac \Sigma V -
1\right)^2 \bar G + \frac {\Sigma - cV} {V^2}. \label{a8a}
\end{eqnarray}
It is our goal in the present paper to establish the two particle
analogues of these weighted Green's functions, i.e. jointly averaged
products of such functions including the
coherent scattering which induces correlations in the joint
 propagation of two particles.

In order to deduce these jointly averaged weighted
functions, we first need to obtain the non-averaged weighted
 functions,
starting with $G^i$ and $G^{ii}$, in a representation  such that no
products between the disorder potential $U$ and the unweighted single
 particle function $G$ occur.
To accomplish this, one can simply employ equation (\ref{a1}) which
 yields
\begin{eqnarray}
G^i & = & \frac {g^{-1} G - 1} V \label{a9} \\
G^{ii} & = & \frac 1 {V^2} \left[ g^{-1}Gg^{-1} - g^{-1} - U \right]
.\label{a10} \\
\end{eqnarray}
and by means of (\ref{a8}) these go over to
\begin{eqnarray}
G^i & = & \frac 1 V \left[ g^{-1} \bar G + g^{-1} \bar G T \bar G -1
\right] \label{a11} \\
G^{ii}&  = & \frac 1 {V^2} \left[g^{-1} \bar G g^{-1} + g^{-1} \bar G
 T \bar G g^{-1} - g^{-1} - U \right]  \label{a12}.
\end{eqnarray}
A further single particle identity resulting from (\ref{a2}) which
will prove to be very useful is
is
\begin{equation}
g^{-1} \bar G = \bar G g^{-1} = \Sigma \bar G + 1 . \label{a13}
\end{equation}

\section{Two particle theory}

In this section a general weighted two particle theory
involving two different bands is established from which the case of
 two particles
moving in a single band can also be immediately obtained.
The calculation for the unweighted two particle function has been made
in {\bf I}. As a main result of that work
 the appropriate vertex corrections for the CPA were obtained
which account for the coherent scattering processes
of two particles that arise in the otherwise
non-interacting two particle function through the averaging process.

We follow the outline of Velick{\'y}'s reasoning to
obtain the proper weights for two choices of the functions $ \langle
 G^{ \mu \nu}_aCG^{\mu^{\prime} \nu^{\prime}}_b \rangle $. The labels
$ \mu, \nu, \mu^{\prime}, \nu^{\prime} \in
\{i,h,0\} $ indicate the kind of weight -- either an impurity, a
host or no weight -- which is attributed to the first and second site
of the respective function. $C$ represents a
generalized  operator coupling the two single particle functions and
$a$ and $b$ label two possibly different bands on which the respective
single particle resolvents are defined. The positions of the
weights within one single Green's function are thereby important,
since the disordered medium before averaging is neither homogeneous
nor isotropic and thus the non-averaged Green's functions depend
non-trivially on both
arguments. In principle there now would be 80 (!)  possible ways of
 applying
specific weights to these functions before averaging. However, in
 most cases, even if different
bands are involved, only two particle functions having an equal
kind of weighting on its single particle constituents will be needed
in most applications.
Later, we will consider two particular examples of two particle
functions which find frequent use in linear response theory where the
 operator $C$ is diagonal in real
space and also only diagonal elements (or sums of diagonal elements)
 of the
functions defined above are used. For both cases considered, the
 total choices of
weightings reduce to only five different ones,
since the first and the second site of the first function will be the
same as the second and first site of the second function,
respectively, which implies that they must also pairwise bear the same
weighting label.

As an example we calculate for a most general choice of $C$ only the
 two functions $ \langle G_a^{ii} C G_b^{ii}
\rangle $ and $ \langle G_a^{i0} C G_b^{0i}
\rangle $ in this section, since they will prove to be the most useful
types of weightings for the
 cases discussed thereafter and all other ones could be obtained in
complete analogy to the calculations presented here.

Using the identities (\ref{a11}) and (\ref{a12}), the single weighted
 two particle function can be written in terms of the single particle
 functions as
\begin{equation}
\langle G^{i0}_a C G^{0i}_b \rangle = V_a^{-1}
V_b^{-1} \langle \left[ g^{-1}_a \bar
G_a + g^{-1}_a \bar
G_a T_a \bar G_a -1 \right] C \left[ \bar G_b T_b \bar G_b g^{-1}_b +
\bar G_b g_b^{-1} -1 \right] \rangle \label{b4}
\end{equation}

and the double weighted one as
\begin{eqnarray}
\langle G^{ii}_aCG^{ii}_b \rangle & = & V_a^{-2} V_b^{-2} \langle
\left[(g_a^{-1} \bar G_a g_a^{-1} -g_a^{-1}) + g_a^{-1} \bar G_a T_a
 \bar G_a
g_a^{-1} - U_a \right]C \nonumber \\
& \times & \left[(g_b^{-1} \bar G_b g_b^{-1} - g_b^{-1}) + g_b^{-1}
\bar G_b T_b \bar G_b g_b^{-1} - U_b \right] \rangle . \label{b5}
\end{eqnarray}
We calculate the double weighted function first, since it
provides the more difficult task and from its solution it is
 straightforward
to derive the one for the single weighted function as well. The
 problem of evaluating
(\ref{b5}) is divided into two parts, the first one
involving all terms not containing the matrix $U$, and the second
 one containing the remainder, i.e.
\begin{equation}
V_a^2V_b^2\langle G^{ii}_aCG^{ii}_b \rangle = {\cal K} + {\cal M}
\label{b6}
\end{equation}
\begin{eqnarray}
{\cal K} & = &
\left[g_a^{-1} \bar G_a g_a^{-1} - g_a^{-1}\right] C \left[g_b^{-1}
\bar G_b g_b^{-1} -g_b^{-1}\right] + g_a^{-1} \bar G_a \langle T_a
\bar G_a g_a^{-1}C g_b^{-1} \bar G_b T_b \rangle \bar G_b g_b^{-1}
\label{b7} \\
{\cal M} & = & \langle U_aCU_b \rangle + \left[ g^{-1}_a C \langle U_b
\rangle - g^{-1}_a \bar G_a g^{-1}_a C \langle U_b \rangle - g^{-1}_a
\bar G_a \langle T_a \bar G_a g^{-1}_a C U_b \rangle + (a
\leftrightarrow b) \right] \label{b8}
\end{eqnarray}
where in ${\cal K}$ the terms involving an average over a single
 T-matrix have
vanished, which is the standard CPA condition and $(a
\leftrightarrow b)$ indicates that the labels are exchanged and the
corresponding expressions reflected around the operator $C$.
We evaluate ${\cal K}$ first since it is the
term needed in the wider range of applications.

With the identity (\ref{a13}) we find that
\begin{equation}
\left[g^{-1}_a \bar G_a g^{-1}_a - g^{-1}_a\right] C \left[g^{-1}_b
\bar G_b g^{-1}_b -g^{-1}_b\right] = \Sigma_a \Sigma_b  g_a^{-1} \bar
 G_a C \bar G_b g^{-1}_b \label{b9}.
\end{equation}

The T-matrix can be decomposed into its single site contributions
$T_n$ as
\begin{equation}
T = \sum_n T_n + \sum_{n \not= m} T_n \bar G T_m + \sum_{n \not= m
\not= l} T_n \bar G T_m \bar G T_l + ... \label{b11}
\end{equation}
Thereby the characteristic exclusions in the sums prevent the particle
from scattering twice in sequence on the same site and $T_n$
satisfies equation (\ref{a3}).

As shown by Velick{\'y}, $T$ can then be replaced in two ways by a
 closed set of equations, namely
\begin{equation}
T = \sum_n Q_n = \sum_n \tilde Q_n \label{b12}
\end{equation}
where
\begin{equation}
Q_n= T_n(1+ \bar G \sum_{n \not= m} Q_m) \label{b13}
\end{equation}
and
\begin{equation}
\tilde Q_n = (1+ \sum_{n \not= m} \tilde Q_m \bar G) T_n.
 \label{b13_5}
\end{equation}
Due to the requirement that $\langle T \rangle =0$ and the single
 site
decomposition of $T$ from (\ref{b11}) it is possible to decompose
averages on different sites
to give
\begin{equation}
0 \equiv \langle T \rangle = \sum_n \langle Q_n \rangle = \sum_n
\langle T_n \rangle \left( 1 + \bar G \sum_{m \not= n} \langle Q_m
\rangle \right) \label{b14}
\end{equation}
This also implies that $\langle Q_n \rangle = \langle \tilde Q_n
 \rangle = 0$. A vertex
function $\Gamma$ can now be defined similar to {\bf I} such that
\begin{equation}
{\cal K}  = g^{-1}_a \bar G_a ( \Sigma_a \Sigma_b C + \Gamma ) \bar
G_b g^{-1}_b \label{b14_5}
\end{equation}
where now
\begin{equation}
\Gamma = \langle T_a \bar G_a g^{-1}_a C g^{-1}_b \bar G_b T_b \rangle
. \label{b15}
\end{equation}
$\Gamma$ can be manipulated along the lines of {\bf I}
by using (\ref{b12}) to yield

\begin{equation}
\Gamma=\sum_n \sum_m \langle Q_n^a \bar G_a g^{-1}_a C g_b^{-1}
 \bar G_b \tilde Q_m^b \rangle \label{b16}.
\end{equation}
By means of (\ref{b13}) - (\ref{b14})  this can be cast into
the form
\begin{equation}
\Gamma_n = \langle T_n^a \bar G_a g^{-1}_a(C + g_a \sum_{p \not= n}
 \Gamma_p g_b)g^{-1}_b \bar G_b T_n^b \rangle \label{b17}
\end{equation}
where now  $\Gamma = \sum_n \Gamma_n$, since from the decoupling
 introduced in (\ref{b14}) one gets $ \langle
Q_n^a \bar G_a g^{-1}_a C g_b^{-1} \bar G_b \tilde Q_m^b \rangle =
\langle Q_n^a \bar G_a g^{-1}_a C g_b^{-1} \bar G_b \tilde Q_n^b
\rangle \delta_{n,m}$.
The only difference to the corresponding expression in {\bf I},
(cf. (47) there) is that $\sum_{p \not= n} \Gamma_p$ is
surrounded by the propagators of the pure medium $g_{a/b}$ here.
We can then use (\ref{b14_5}) and (\ref{b17}) to obtain
\begin{equation}
\Gamma_n = \langle T_n^a \bar G_a g^{-1}_a C g^{-1}_b \bar G_b
 T_n^b \rangle + \langle T_n^a g_a{\cal K}g_b T_n^b \rangle -
\Sigma_a \Sigma_b \langle T_n^a \bar G_a C \bar G_b T_n^b \rangle
 - \langle T_n^a \bar G_a \Gamma_n \bar G_b T_n ^b \rangle
 \label{b19}.
\end{equation}

At this stage we are able to find $\Gamma_n$ and therefore also
 ${\cal K}$.
As we cast our model into a site representation we obtain $T_n =
 \mid n \rangle t_n \langle n \mid$ and hence
\begin{equation}
\Gamma_n=\mid n \rangle \Lambda \langle n \mid \left[ g_a{\cal K}g_b
 + \bar G_a g^{-1}_a C g^{-1}_b \bar G_b - \Sigma_a \Sigma_b \bar
 G_a C \bar G_b \right] \mid n \rangle \langle n \mid \label{b20}
\end{equation}
where $\Lambda$ is the irreducible vertex part derived by
Velick{\'y}
\begin{equation}
\Lambda(z_1,z_2) = \frac {\langle t_n^a(z_1) t_n^b(z_2) \rangle}
 {1+ F_a(z_1) \langle t_n^a(z_1) t_n^b(z_2) \rangle F_b(z_2) }
 \label{b21}.
\end{equation}
 The vertex $\Lambda$ can be regarded as being intrinsic to the CPA,
since it does not depend on the particular form of the operator $C$.
Substituting (\ref{b20}) into (\ref{b14_5}) yields
\begin{eqnarray}
& &{\cal K}= \Sigma_a \Sigma_b g^{-1}_a \bar G_a C \bar G_b
  g^{-1}_b +
\nonumber \\
& + & \Lambda g^{-1}_a \bar G_a \sum_n \mid n \rangle  \langle n
 \mid \left[ g_a{\cal K}g_b + \bar G_a g^{-1}_a C g^{-1}_b \bar G_b
 - \Sigma_a \Sigma_b \bar G_a C \bar G_b \right] \mid n \rangle
\langle n \mid \bar G_b  g^{-1}_b. \label{b22}
\end{eqnarray}
Multiplying by $g_a$ and $g_b$ from the left and right, we solve for
the diagonal elements $ \langle m \mid g_a{\cal K}g_b \mid m \rangle
 $
\[
\langle m \mid g_a{\cal K}g_b \mid m \rangle = \Sigma_a \Sigma_b
 \langle m \mid \bar G_a C \bar G_b \mid m \rangle +
\]
\begin{equation}
+ \Lambda \sum_n  \langle m \mid \bar G_a \mid n \rangle  \langle n
 \mid \left[ g_a{\cal K}g_b + \bar G_a g^{-1}_a C g^{-1}_b \bar G_b
 - \Sigma_a \Sigma_b \bar G_a C \bar G_b \right] \mid n \rangle
\langle n \mid \bar G_b \mid m \rangle \label{b23}
\end{equation}
and hence also solve (\ref{b22}).
At this point it is helpful to visualize the form of
the operator $C$ in the site representation, which in the most general
case can be written as
\begin{equation}
C = \sum_{l,m} \gamma_{lm} \mid l \rangle \langle m \mid \label{b24}
\end{equation}
For convenience we introduce the short notation
\begin{eqnarray}
F_{n-m} & = & \langle n \mid \bar G \mid m \rangle ,\label{b25} \\
{\cal F}_{n-m} & = & \langle n \mid g^{-1} \bar G \mid m \rangle
 \label{b26}
\end{eqnarray}
where $F_0 \equiv F$ as already defined in (\ref{a3a}).
It should be noted that $F_{n-m}$ and ${\cal F}_{n-m}$ are of
different dimensions and their definition has been chosen to reduce
 the
algebra as much as possible. Furthermore, these expressions show
that equation (\ref{b23}) only contains translationally
invariant quantities and hence it can be solved by Fourier
 transformation.
The following Fourier transforms are introduced
\begin{eqnarray}
a_k & = & \Sigma_a \Sigma_b \sum_m e^{-ikR_m} \langle m \mid \bar G_a
 C \bar G_b \mid m \rangle \label{b28} \\
\alpha_k  & =  & \sum_m e^{-ikR_m} \langle m \mid g_a^{-1} \bar G_a C
 \bar G_b
g^{-1}_b \mid m \rangle = \nonumber \\
& =  & \sum_m e^{-ikR_m} \left[ \Sigma_a \Sigma_b \langle m
\mid \bar G_a C \bar G_b \mid m \rangle + \gamma_{mm} + \Sigma_a
 \sum_n
\gamma_{nm} F^a_{m-n} + \Sigma_b \sum_n \gamma_{mn} F^b_{n-m} \right]
\label{b32} \\
A_k  & = & \sum_m e^{-ikR_m} F^a_m F^b_{-m} \label{b30} \\
{\cal A}_k  & =  & \sum_m e^{-ikR_m} {\cal F}^a_m {\cal F}^b_{-m}
\label{b31} \\
b_k  & =  & \sum_m e^{-ikR_m} \langle m \mid g_a{\cal K}g_b \mid m
 \rangle
\label{b33} \\
c_k  & =  & \sum_m e^{-ikR_m} \langle m \mid {\cal K} \mid m \rangle
 \label{b34}.
\end{eqnarray}
The units of $a_k, \alpha_k, b_k, {\cal A}_k$ are unity, that of
$A_k$ is $J^{-2}$ and that of $c_k$ is $J^2$.
Inserting into (\ref{b23}) yields
\begin{equation}
b_k = \frac {a_k + \Lambda A_k (\alpha_k - a_k)} {1 - \Lambda A_k}
 \label{b34_5},
\end{equation}
and hence by means of (\ref{b22})
\begin{equation}
c_k  = \alpha_k \frac {\Sigma_a \Sigma_b + \Lambda ( {\cal A}_k -
 \Sigma_a
\Sigma_b A_k ) } {1 - \Lambda A_k} = \alpha_k \frac {\Sigma_a
 \Sigma_b + \Lambda D_k } {1 - \Lambda A_k}  \label{b35}
\end{equation}
where
\begin{equation}
D_k = {\cal A}_k - \Sigma_a \Sigma_b A_k = \sum_m e^{-ikR_m} \left[
\gamma_{mm} + \Sigma_a \sum_n
\gamma_{nm} F^a_{m-n} + \Sigma_b \sum_n \gamma_{mn} F^b_{n-m}
 \right] \label{b37}.
\end{equation}
Similarly, one can evaluate the term ${\cal M}$ from (\ref{b8}),
 but it
turns out that little simplification can be made until an explicit
form of $C$ is known and therefore we postpone its evaluation to the
 next section.

For now, we proceed to calculate the single weighted function $\langle
G^{i0}_a C G^{0i}_b \rangle$. From (\ref{b4}) we get
\begin{equation}
V_a V_b \langle G^{i0}_a C G^{0i}_b \rangle = g^{-1}_a \bar G_a C
 \bar G_b
g^{-1}_b + C - g^{-1}_a \bar G_a C - C \bar G_b g^{-1}_b + g^{-1}_a
 \bar G_a
\langle T_a \bar G_a C \bar G_b T_b \rangle \bar G_b g^{-1}_b.
 \label{bb1}
\end{equation}
Using (\ref{a13}), this  can be recast into
\begin{equation}
V_a V_b \langle G^{i0}_a C G^{0i}_b \rangle = \Sigma_a \Sigma_b \bar
 G_a C \bar G_b + g^{-1}_a \bar G_a
\langle T_a \bar G_a C \bar G_b T_b \rangle \bar G_b g^{-1}_b
 \label{bb2}
\end{equation}
which can be readily solved since the term $\langle
T_a \bar G_a C \bar G_b T_b \rangle$ exactly corresponds to
the vertex part of the unweighted function which is known from {\bf I}
\begin{equation}
\langle T_a \bar G_a C \bar G_b T_b \rangle = \sum_n \mid n \rangle
\Lambda \langle n \mid \langle G_a C G_b \rangle \mid n \rangle
 \langle n
\mid \label{bb2.5}.
\end{equation}
 Thus, the Fourier transform
\begin{equation}
d_k = V_a V_b \sum_m e^{ikR_m} \langle m \mid  \langle G^{i0}_a C
 G^{0i}_b \rangle \mid m
\rangle \label{bb3}
\end{equation}
can be  obtained using the solution for the unweighted two
particle function and equations (\ref{b28}), (\ref{b30}) and
(\ref{b31}) to yield
\begin{equation}
d_k = a_k + \Lambda {\cal A}_k \frac {a_k} { \Sigma_a \Sigma_b (1 -
\Lambda A_k)} \label{bb4}.
\end{equation}
Using (\ref{b37}) this can be written as
\begin{equation}
d_k = \frac {a_k} {\Sigma_a \Sigma_b} \frac {\Sigma_a \Sigma_b +
 \Lambda D_k} {1 -
\Lambda A_k }. \label{bb5}
\end{equation}
It is important to indicate at this point that the solution for
a single weighted function in which the weights have been swapped
 to the inside, i.e. $
\langle G^{0i}_a C G^{i0}_b \rangle $, will not be the same as the one
just obtained. A similar type of calculation yields instead of
(\ref{bb4})
\begin{equation}
d_k^{ \prime } = a_k + \Lambda A_k \frac {\alpha_k} {1 -
\Lambda A_k}, \label{bb6}
\end{equation}
where $ d_k^{ \prime } = \sum_n e^{ikR_m} \langle m \mid
\langle G^{0i}_a C G^{i0}_b \rangle \mid m \rangle $.

Since we have now obtained expressions for the Fourier transformed
 site diagonal
elements of some of the impurity weighted two particle functions, we
are also able to find their respective off diagonal parts by
 inserting into
the appropriate Bethe-Salpeter equations ((\ref{b22}) in the case of
 the double weighted function). Other types of weighted functions
 can be
calculated from here by setting up the
corresponding equations in analogy to (\ref{b4}) and (\ref{b5}) from
the single particle functions. In particular, to obtain the
 corresponding host
weighted functions one can apply the weights $(1 - U_{a/b}/V_{a/b})$
 to the
single particle resolvents from either side, but the form of this
weight already suggests that host weighted properties can always be
deduced by adding and/or subtracting corresponding
 unweighted/impurity weighted functions. For the most general case
 the set of
closed equations will be quite large and for that reason we
will in the following deduce these equations only in some more
 specific cases.

\section{Applications in linear response theory}

The most obvious application of a two particle theory
as introduced above is in linear response theory, i.e. in calculating
weighted susceptibilities. The study of weighted functions in this
context helps to
determine how such quantities as susceptibilities and transport
coefficients are constituted (on the average) from processes on
different components of the alloy. A further advantage of such a
differentiation is that it allows for a more refined treatment of
further renormalizations to the considered quantities once
further interactions are introduced into the problem.
The standard expression employed in linear response theory is a
 generalized Kubo formula
\begin{equation}
\chi_{C^{(1)},C^{(2)}}(z) = \int d \xi d \eta S_0 ( z,\xi,\eta )
Z^{-1} Tr_b \{ e^{- \beta H } C^{\dagger (2)} \langle \hat
 \delta^{\mu \nu} (\xi - H_a  ) C^{(1)} \hat \delta^{ \mu \nu }
  (\eta - H_b ) \rangle \} . \label{c1}
\end{equation}
Thereby $ \chi_{C^{(1)},C^{(2)}} $ stands for a generalized type
 of susceptibility characterizing the linear response of an
 observable $C^{(1)}$ to an external perturbation coupling into
 the Hamiltonian through the operator  $C^{(2)}$.
$S_0$ denotes the zero order associated two particle Matsubara
 function in
the Lehmann representation \cite{bi25} and $\hat \delta^{
\mu \nu} (\xi - H_{a/b}) $ is the corresponding spectral function
of the full
resolvent $ G^{ \mu \nu}_{a/b}  ( \xi ) = \Xi^{\mu}_{a/b}
 [ \xi - H_{a/b} ]^{-1} \Xi^{\nu}_{a/b}$, defined on the
 subspace (band) $ a/b $, with
the weighting operators $\Xi^{\mu}_{a/b}$ applied to it
 which assume the values $\Xi^i_{a/b} = U_{a/b}/V_{a/b}$,
 $\Xi^h_{a/b} = (1-U_{a/b}/V_{a/b})$ and $\Xi^0_{a/b} = 1$. $Tr_b$
denotes the trace over the subspace (band) $b$.
Our particular interest focuses on the operators $C^{(1)}$ and
 $C^{(2)}$ which we now take to have the following diagonal form
\begin{equation}
C^{(1)/(2)} = \sum_m \gamma_m^{(1)/(2)} (\xi, \eta) \mid m
  \rangle \langle m  \mid \label{c2a}.
\end{equation}
To examine the behavior of the weighted functions further under such a
constraint it is useful to
consider the two cases:
\begin{equation}
 \gamma_m = const \times \delta_{m,r} \label{c2}
\end{equation}
 where $r$ is an explicitly specified site and
\begin{equation}
 \gamma_m = const . \label{c3}
\end{equation}
The first case arises in a treatment of the response from local
 interactions such
as for example the space dependent exchange
interaction between spin impurities
(RKKY interaction \cite{bi21})
submerged in a system otherwise containing non-magnetic disorder
\cite{bi22}. The second case is widely
used in calculations of the optical response in metals and
 semiconductors\cite{bi15,bi20}, since the exciting optical
fields can be taken as uniform over space with approximate momentum
$q=0$ and the optical matrix elements
describing transitions between bands of different angular momentum
symmetry are usually well approximated as constants.

In both cases the restriction to  only consider diagonal elements of
 two
particle functions reduces the set of closed equations for all
 possible
weightings to
\begin{equation}
\langle G^h_a C G^h_b \rangle = \langle G_a C G_b \rangle -
 \langle G^i_a C G^i_b
 \rangle \label{c4}
\end{equation}
\begin{equation}
\langle G^{ih}_a C G^{ih}_b \rangle = \langle G^{hi}_a C G^{hi}_b
 \rangle
= \langle G^i_a C G^i _b\rangle - \langle G^{ii}_a C G^{ii}_b
 \rangle \label{c5}
\end{equation}
\begin{equation}
\langle G^{hh}_a C G^{hh}_b \rangle = \langle G^h_a C G^h_b \rangle
 - \langle
G^{ih}_a C G^{ih}_b \rangle  \label{c6}
\end{equation}
It follows, that for this case the calculation of only $ \langle
G^{ii}_aCG^{ii}_b \rangle $ and $ \langle G^i_aCG^i_b \rangle $ is
sufficient to also obtain the remaining mixed and host
weighted functions.
We start to consider the case of (\ref{c2}) first.

\subsection{Electronic Susceptibility}

The response of a system at some point $r_1$ in space to a
perturbation applied at point $r_2$ has a
wide range of applicability. Disregarding the coupling constants which
simply scale the result we assume that $C^{(1)}  = \mid 2 \rangle
\langle 2 \mid $ and $C^{(2)}  = \mid 1 \rangle
\langle 1 \mid $, where the choice of the origin is arbitrary. In a
slightly different notation the calculation of the double weighted
 function
corresponds to evaluating
$\langle  G_a^{ii}(r_1,r_2;t) G_b^{ii}(r_2,r_1;t^{ \prime }) \rangle$.

For the RKKY interaction mentioned before, which couples spins at
 a distance
$r_1 - r_2$ through the electronic spin susceptibility of
electrons in the conduction band, the two-particle time
dependent function is to be Fourier transformed and to be taken at two
identical single particle energies
\cite{bi21}. If the spins are at impurity sites, the weighted
functions must be used to describe the problem adequately. Earlier
treatments \cite{bi22} neglected the vertex corrections and
it is proposed to investigate the effect of their inclusion in a
separate publication.

In this as in other problems the self-interaction $r_1=r_2$ can
be neglected for most purposes, which implies that it will be
sufficient to calculate only the
term ${\cal K}$ from (\ref{b7}), since all terms occurring in
${\cal M}$ from (\ref{b8}) will vanish for $r_1 \not= r_2$, since
 the disorder
potential $U$ is a diagonal matrix which vanishes identically for
 off-diagonal terms.
To clarify the meaning of the quantity which is obtained through this
special choice of $C^{(1)}$ and $C^{(2)}$, we look at how the
corresponding Bethe-Salpeter (B-S) equation can be rewritten for
 the case of a unweighted two particle Green's function as
introduced in equations (22) of Ref. \onlinecite{bi8}. In the single
 site approximation the B-S equation for the unweighted function
 reads
\begin{eqnarray}
\langle G_a(1,2) G_b(1,2)  \rangle & \equiv & \langle
G_{ab}^{(2)}(1,1;2,2) \rangle =  \langle G_a(1,2) \rangle \langle
G_b(1,2) \rangle + \nonumber \\
& + &  \Lambda \sum_n \langle G_a(1,n) \rangle \langle G_b(1,n)
 \rangle
\langle G_{ab}^{(2)}(n,n;2,2) \rangle \label{ca1}
\end{eqnarray}
where we have rewritten
\begin{equation}
\langle 2 \mid \bar G_a \mid 1 \rangle \langle 1 \mid \bar G_b \mid 2
\rangle \equiv  \bar G_a(1,2) \bar G_b(1,2)  \label{ca2}
\end{equation}
and
\begin{equation}
\langle 2 \mid \langle G_a \mid 1 \rangle \langle 1 \mid G_b \rangle
 \mid 2 \rangle \equiv
\langle G_{ab}^{(2)}(1,1;2,2) \rangle \label{ca3}.
\end{equation}
For the Fourier transforms from before (\ref{b28}) - (\ref{b31}) this
amounts to
\begin{eqnarray}
a_k & = & \Sigma_a \Sigma_b \sum_m e^{-ikR_m} \bar G_a(m) \bar G_b(m)
\label{ca4} \\
\frac {a_k} {\Sigma_a \Sigma_b} & = & A_k \label{ca5} \\
\alpha_k & = & \sum_m e^{-ikR_m} \langle m \mid g_a^{-1} \bar G_a
 \mid 1 \rangle \langle 1 \mid \bar G_b
g_b^{-1} \mid m \rangle = \nonumber \\
& = & \sum_m e^{-ikR_m} \left[ \Sigma_a \Sigma_b \bar G_a(m) \bar
 G_b(m) +
\delta_{m,1} (1 +  \Sigma_a F_a + \Sigma_b F_b)
\right]  = {\cal A}_k .\label{ca6}
\end{eqnarray}
Using (\ref{ca5}) and (\ref{ca6}) in connection
with (\ref{bb4}) and (\ref{bb6}) the
result for the single weighted functions is independent of
whether the impurity weights are applied to the interior or to the
exterior of the single particle resolvents, such that $ \langle
 G_a^{i0}(1,2)
G_b^{0i}(1,2)\rangle  =  \langle G_a^{0i}(1,2) G_b^{i0}(1,2) \rangle
$. Indeed it is not hard to show that other arbitrary distributions
of two impurity weights also give the same results as long as there
 is one
weight applied to each resolvent. The same is of course
true for host related properties.

Accordingly,  $b_k$ from (\ref{b34_5}) in the previous section goes
 over to
\begin{equation}
b_k = a_k \frac {\Sigma_a \Sigma_b + \Lambda (\alpha_k - a_k)}
{\Sigma_a \Sigma_b - \Lambda a_k} \label{ca7}
\end{equation}
and similarly $c_k$ from equation (\ref{b35}) to
\begin{equation}
c_k = \Sigma_a \Sigma_b
\alpha_k \left( 1+\Lambda \frac {\alpha_k} {\Sigma_a \Sigma_b -
\Lambda a_k} \right) \label{ca8}.
\end{equation}
We remember that the Fourier transform of the unweighted two particle
Greens function is
\begin{equation}
\sum_{R_{1,2}} e^{ikR_{1,2}} \langle  G_a(1,2) G_b(1,2)  \rangle =
\frac {a_k} {\Sigma_a \Sigma_b - \Lambda a_k } . \label{ca9}
\end{equation}

Phenomenologically, one can scrutinize the uncorrelated limit where $
\Lambda \rightarrow 0$ for which in (\ref{ca8}) the $k$- dependent
 part of $c_k$ behaves as
\begin{equation}
\lim_{\Lambda \rightarrow 0} c_k = \Sigma_a \Sigma_b a_k + ...
 \label{ca10}
\end{equation}
which is the correct limiting result for a product of two
averaged double impurity weighted single particle functions
$ \bar G_a^{ii}(1,2) \bar
G_b^{ii}(1,2)$ or the weighted two particle function without coherent
corrections.

However, taking this limit is reasonable only in special cases since
generally the vertex $\Lambda$ depends on the self energy $\Sigma$. It
will be appropriate to take in a weak disorder limit (virtual crystal
limit) since for $V_{a/b} \rightarrow 0$ the CPA predicts
that $\Sigma_{a/b} \rightarrow c V_{a/b}$ and $\Lambda \rightarrow
c (1-c)
V_a V_b$ which means that $\Lambda$ approaches zero faster than
$\Sigma_{a/b}$ in this limit.

One should also note at this point, that although it is apparent that
 the
term $(1+ \Sigma_a F_a + \Sigma_b F_b)$ , which arises in the
 difference of the
Fourier transforms $\alpha_k$ and $a_k$, originates from
the difference in the diagonal parts of their respective real space
Green's functions, it can not be neglected in this
treatment. When $\Lambda$ is finite, this term
is multiplied with other $k-$dependent quantities and thus
 contributes to the off diagonal elements as well.
Written in terms of $a_k$, (\ref{ca8}) can be cast into
\begin{equation}
c_k = \frac {a_k} { \Sigma_a \Sigma_b - \Lambda a_k} \left(
\Sigma_a \Sigma_b + \Sigma_a \Sigma_b (a_k)^{-1} (1 + \Sigma_a F_a +
 \Sigma_b F_b) \right)
\left( \Sigma_a \Sigma_b + \Lambda (1 +  \Sigma_a F_a + \Sigma_a F_b)
 \right) \label{ca11}.
\end{equation}
If again diagonal terms in real space are neglected,
i.e. $k-$independent quantities in $k-$space, it can be shown that
(\ref{ca11}) can be cast into the compact form
\begin{equation}
\zeta_k = \frac {a_k} {\Sigma_a \Sigma_b - \Lambda a_k} {\left(
\Sigma_a \Sigma_b + \Lambda (1 + \Sigma_a F_a + \Sigma_b F_b)
 \right)}^2 \label{ca12}
\end{equation}
where we have introduced $\zeta_k \equiv c_k + c_o$, whereby $c_o$ is
independent of $k$.
According to the definition of $\zeta_k$, this finally relates the
 weighted off-diagonal real space two particle Green's function to
 the unweighted one as
\begin{equation}
\langle G_a^{ii}(1,2) G_b^{ii}(1,2)  \rangle = \langle G_a(1,2)
G_b(1,2) \rangle \frac {{\left( \Sigma_a \Sigma_b + \Lambda (1 +
\Sigma_a F_a + \Sigma_b F_b) \right)}^2} {(V_a V_b)^2} . \label{ca13}
\end{equation}

Equally, the relation for the single weighted function $\langle
 G_a^i(1,2) G_b^i(1,2)\rangle$ can be obtained almost immediately
if (\ref{bb4}) is modified for this choice of $C^{(1)}$ and $C^{(2)}$,
which goes over to
\begin{equation}
d_k = a_k + \Lambda \alpha_k \frac {a_k} { \Sigma_a \Sigma_b  -
\Lambda a_k} = \frac {a_k} { \Sigma_a \Sigma_b - \Lambda a_k} \left[
\Sigma_a \Sigma_b + \Lambda (1 + \Sigma_a F_a + \Sigma_b F_b ) \right]
 \label{ca14}
\end{equation}
thus relating the real space functions in this case as
\begin{equation}
\langle G_a^i(1,2) G_b^i(1,2) \rangle = \langle G_a(1,2) G_b(1,2)
 \rangle
\frac { \Sigma_a \Sigma_b + \Lambda (1 + \Sigma_a F_a \Sigma_b F_b )}
{V_a V_b} \label{ca15}
\end{equation}
where no diagonal contributions were omitted.

The fact that in the single weighted case the same weighting
factor occurs once as opposed to twice for the double weighted
one is structurally equivalent to the results for the single
particle theory. Although retrospectively that might not be
 surprising,
it is also quite interesting in view of the fact that for the most
general case of the previous section the weighting factors in
(\ref{b35}) and (\ref{bb5}) almost look alike were it not for the
difference in the pre-factors, i.e. $\alpha_k$ in the former and
$a_k / \Sigma_a \Sigma_b$ in the latter case.

In principle all other functions can be derived now using the set of
relations (\ref{c4}) - (\ref{c6}).
However, one can save a considerable amount of algebra by recalling
 the following relation which holds in the transition from impurity
 to host related properties in the CPA
\begin{equation}
- \Sigma \rightarrow (V - \Sigma) \hspace{20mm} c \rightarrow (1-c).
 \label{ca16}
\end{equation}
Starting from the definition of the vertex $ \Lambda $ in
(\ref{b21}) and the CPA condition $ \langle T_n \rangle  = 0 $ one
finds that
\begin{equation}
\Lambda = \left[ \frac {1-c} {(V_a - \Sigma_a)(V_b - \Sigma_b)} +
 \frac
c { \Sigma_a \Sigma_b } \right]^{-1} \label{ca17}
\end{equation}
 which in a single band case ($a=b$) readily simplifies to the form
 first
introduced by Leath \cite{bi8}
\begin{equation}
\Lambda_{a=b} = \frac {\delta \Sigma} {\delta F} = \frac { \Sigma
( V - \Sigma ) } { 1- ( V - 2 \Sigma) F }. \label{ca18}
\end{equation}
With the relation (\ref{a3a}) between $F$ and $\Sigma$
the weighting factor for the single weighted impurity function from
(\ref{ca15}) reduces to
\begin{equation}
\frac { \Sigma_a \Sigma_b + \Lambda (1 +  \Sigma_a F_a  +
\Sigma_b F_b) }
{V_a V_b} =  \frac {(1-c) \Lambda}
 {(V_a - \Sigma_a)(V_b - \Sigma_b)}. \label{ca19}
\end{equation}
Correspondingly, the factor for the double weighted function is  the
square of this quantity.
One can now employ (\ref{c4}) and (\ref{c5}) to find the weights
 for the single host- and the impurity-host weighted functions,
 respectively.
We find
\begin{equation}
\langle G_a^h G_b^h  \rangle =  \langle G_a G_b \rangle \frac {c
\Lambda^2} {\Sigma_a \Sigma_b} \label{ca20}
\end{equation}
and
\begin{equation}
\langle G_a^{ih}  G_b^{ih} \rangle =  \langle G_a G_b  \rangle \frac
{c
(1-c)\Lambda} {\Sigma_a \Sigma_b (V_a - \Sigma_a ) (V_b - \Sigma_b )}
 \label{ca21}
\end{equation}
which by means of (\ref{c6}) gives
\begin{equation}
\langle G_a^{hh}  G_b^{hh}\rangle = \langle G_a G_b \rangle
\left( \frac {c \Lambda} { \Sigma_a \Sigma_b } \right)^2. \label{ca22}
\end{equation}
Note that this could also have been expected from the transformation
 property
(\ref{ca16}).
The impurity and host weights as represented in (\ref{ca19}) and
(\ref{ca20}) are immediately seen to be the two contributions summing
to $ \Lambda^{-1} $  in (\ref{ca17}). Thus, even though there
are many different possible representations of the two band vertex,
the representation in (\ref{ca17}) shows that the transition from
 impurity to host
properties leaves $\Lambda$ invariant.

A notable feature about the weights calculated in this
section is that they are independent of any wavevectors and only
multiply the unweighted function as scalar energy
dependent factors.
This is a direct consequence of the single site
approximation.

\subsection{Theory of Absorption}

In this section we evaluate a form of the two particle functions
needed in the calculation of the linear response absorption in a
disordered solid. We take the
operators $C^{(1)}$ and $C^{(2)}$ in the characteristic form of
 dipole
operators similar to the one in (\ref{c3}). Furthermore, we assume
 that the dipole matrix elements be
essentially constant, such that $ C^{(1)/(2)}= \gamma^{(1)/(2)} \sum_m
\mid m  \rangle \langle m  \mid $. This choice corresponds to
applications for the description of processes involving transitions
 between bands of different angular
momentum symmetry such as required by the selection rule for optical
processes at at zero total momentum
\cite{bi8,bi10}.
At the end of this sub-section we will also
 give for completeness a short account of
the calculation of linear response conductivities in disordered
 solids.

For the above choice of $\gamma^{(1)/(2)}$ the calculation of the
 term ${\cal K}$
in (\ref{b22}) is greatly simplified compared to before, but it will
be necessary now to also consider total diagonal terms, since the
 sums over all
states in $C^{(1)/(2)}$ couple all sites and hence all
contributions coming from term ${\cal M}$ in (\ref{b8}) have to be
included.

As a consequence of the
introduction of the dipole operator the main change arising in the
result for ${\cal K}$ is that the site diagonal elements
$\langle n \mid {\cal K} \mid n \rangle $ and $\langle n \mid
G_a  C^{(1)} G_b \mid n \rangle$  as well as $\langle n \mid
G_a g_a^{-1} C^{(1)} g_b^{-1} G_b \mid n \rangle $ are now actually
 independent
of $n$ \cite{bi15}. Since $C^{(1)}$ now couples the functions to its
 left and right
like a matrix product, the B-S equations  (\ref{b22}) and (\ref{b23})
have a very simple solution in terms of their Fourier transforms.
 Introducing $a,b,c \equiv a_k,
b_k, c_k \mid_{k=0}$ as the zero momentum elements of the
respective transforms from last section we effectively get $c= \langle
n \mid {\cal
K}\mid n \rangle $ , $a =
\Sigma_a \Sigma_b \langle n \mid \bar G_a \bar G_b \mid n \rangle $,
 where the omission
of $C^{(1)}$ indicates that the two single particle resolvents are now
simply multiplied as matrices, and equation (\ref{b34_5}) reduces to
\begin{equation}
b=a \frac {\Sigma_a \Sigma_b + \Lambda (\alpha - a)} {\Sigma_a
\Sigma_b - \Lambda a} \label{cb1}
\end{equation}
and equation (\ref{b35}) to
\begin{equation}
 c = \Sigma_a \Sigma_b \alpha \left( 1 + \Lambda \frac \alpha
 {\Sigma_a
\Sigma_b - \Lambda a} \right) \label{cb2}
\end{equation}
which can be recast into
\begin{eqnarray}
c & = & \frac  a {\Sigma_a \Sigma_b - \Lambda a} \left[ \Sigma_a
\Sigma_b + \Lambda (1 + \Sigma_a F_a + \Sigma_b F_b) \right]^2
\nonumber \\
& + & (1 + \Sigma_a F_a + \Sigma_b F_b) \left[ \Sigma_a
\Sigma_b + \Lambda (1 + \Sigma_a F_a + \Sigma_b F_b) \right]
 \label{cb3}
\end{eqnarray}
where the term independent of $a$ which was discarded in the previous
section has to be kept in this case since the contributions of the
diagonal elements become important.

In calculating ${\cal M}$ one has to be aware that $U$ is a
matrix which just has a random occupation of its diagonal. The sum
over all sites in the operators $C^{(1)}$ and $C^{(2)}$ will hence
 just
pick out the sum of all total diagonal parts $ \langle n \mid
{\cal G} \mid n \rangle \langle n \mid U \mid n \rangle$ where
${\cal G}$ is a generalized product of several Green's functions
 in the same band (the case where $U$ and ${\cal G}$ are swapped
is analogous). From (\ref{b8}) we get immediately
\begin{equation}
{\cal M} = c V_a V_b + \left[ cg_a^{-1}V_b - c g_a^{-1} \bar G_a
g_a^{-1} V_b - g_a^{-1} \bar G_a \langle T_a \bar G_a g_a^{-1} U_b
\rangle + (a \leftrightarrow b) \right] \label{cc1}
\end{equation}
where the only term giving slight complications is $\langle T_a \bar
G_a g_a^{-1} U_b \rangle$. However by means of (\ref{a13}) we obtain
\begin{equation}
\langle T_a \bar G_a g_a^{-1} U_b \rangle = \Sigma_a \langle T_a \bar
 G_a U_b \rangle + \langle T_a U_b \rangle \label{cc2}.
\end{equation}
In this expression, the second term
presents more complications, since for the first one we remember from
 (\ref{a8}) that
\begin{equation}
T \bar G = \bar G^{-1} G - 1 \label{cc3}
\end{equation}
and $U_b = U_a V_b/V_a$ such that
\begin{equation}
\langle T_a \bar G_a U_b \rangle = \langle \bar G_a^{-1} G_a U_b
\rangle - c V_b = \frac { \Sigma_a V_b } {V_a} -cV_b. \label{cc4}
\end{equation}
Subsequently, the second term can be decoupled by means of
(\ref{b12}) -
(\ref{b13_5})
\begin{equation}
\langle T_a U_b \rangle = \sum_n \langle \tilde  Q_n^a U_b \rangle =
\sum_n \langle (1 + \sum_{m \not= n} \tilde Q_m^a \bar G_a ) T_n^a U_b
\rangle \label{cc5},
\end{equation}
Applying (\ref{b14}) and using the the fact
that $U$ is diagonal yields
\begin{equation}
\langle \tilde Q_m^a \bar G_a T_n^a U_b \rangle = \langle \tilde Q_m^a
\rangle \bar G_a \langle T_n^a U_b \rangle = 0 \label{cc6}
\end{equation}
since $(m \not= n)$ and $\langle \tilde Q_m \rangle = 0$ and hence
 we find
\begin{equation}
\langle T_a U_b \rangle = \sum_n \langle T_n^a U_n^b \rangle =
  \frac {c (V_a-
\Sigma_a)V_b} {1- (V_a-
\Sigma_a) F_a}. \label{cc7}
\end{equation}
Collecting all terms for ${\cal M}$ and some
more algebraic manipulation finally yields
\begin{equation}
{\cal M} = cV_a V_b - (1+ \Sigma_a F_a) \Sigma_a V_b - (1+ \Sigma_b
F_b) \Sigma_b V_a. \label{cc8}
\end{equation}

The sum of all diagonal parts, i.e. ${\cal M}$ and the ones from the
 second term in (\ref{cb3}), can be shown to assume the very compact
 form
\begin{equation}
{\cal M} +  (1 + \Sigma_a F_a + \Sigma_b F_b) \left[ \Sigma_a
\Sigma_b + \Lambda (1 + \Sigma_a F_a + \Sigma_b F_b) \right] = \frac
 {\Lambda F_a F_b} { V_a V_b}. \label{cc9}
\end{equation}

The final result for $ \langle G_a^{ii} G_b^{ii} \rangle$ thus amounts
to
\begin{equation}
\langle G_a^{ii} G_b^{ii} \rangle = \langle G_a G_b \rangle \left[
 \frac {(1-c) \Lambda} {(V_a - \Sigma_a )( V_b - \Sigma_b)} \right]^2
 +  \frac {\Lambda F_a F_b} { V_a V_b}. \label{cc10}
\end{equation}
Here the first term has been
rewritten in the same way as already derived in the last subsection
for the finite range susceptibility.
{}From there it is also seen that the single weighted function will have
the same weight as calculated in (\ref{ca19}) for the
corresponding function in the exchange coupling case.
We find
\begin{equation}
\langle G_a^i G_b^i \rangle = \langle G_a G_b \rangle  \frac {(1-c)
 \Lambda} {(V_a - \Sigma_a )( V_b - \Sigma_b)}   \label{cc11}
\end{equation}
and by means of (\ref{c4}) - (\ref{c6})
\begin{equation}
\langle G_a^h G_b^h \rangle = \langle G_a G_b \rangle  \frac {c
\Lambda} {\Sigma_a \Sigma_b} \label{cc12}
\end{equation}
\begin{equation}
\langle G_a^{ih} G_b^{ih} \rangle = \langle G_a G_b \rangle \frac
 {c(1-c) \Lambda^2} {(V_a - \Sigma_a )( V_b - \Sigma_b) \Sigma_a
 \Sigma_b} - \frac {\Lambda F_a F_b} { V_a V_b} \label{cc13}
\end{equation}
\begin{equation}
\langle G_a^{hh} G_b^{hh} \rangle = \langle G_a G_b \rangle \left[
 \frac {c \Lambda} { \Sigma_a  \Sigma_b} \right]^2 +  \frac {\Lambda
 F_a F_b} { V_a V_b}. \label{cc14}
\end{equation}

Comparing the results of the last two sub-sections, it becomes clear
 that the several
weights obtained are essentially universal. The main difference in the
absorption case as compared to the susceptibility one comes from the
 diagonal terms which have to be kept in the
double weighted functions. The single weighted analogues are void of
 this difficulty and
the weighting factors are identical for both cases.

So far we had omitted to consider a form of the two particle functions
which is needed for conductivity calculations. However, Velick{\'y}
\cite{bi4}
showed for the unweighted functions that the vertex $\Gamma$ vanishes
 in the corresponding expression for the conductivity, due to the
 antisymmetry of the dipole
matrix elements in $k$-space if they are taken between Bloch states of
a non interacting Fermi system in
a crystal with inversion symmetry. The same is also true for the
 weighted
case and effectively the weighted functions which would have to be
used for such calculations would just consist of products of the
corresponding single particle quantities.

This however turns out to be a general deficiency of the CPA in the
single site approximation since due to the multiple scattering
exclusions, only ladders of nested diagrams are used in
calculating the total contribution of the coherent
scattering.  The CPA therefore neglects higher order two particle
correlations which are in fact non-zero and contribute markedly
 to the conductivity. Langer
and Neal \cite{bi11} have shown that the so called ``maximally
crossed'' diagrams, i.e. diagrams which have a maximal crossing of
coherent particle-particle scattering lines
actually contribute the leading part -- in the order of the expansion
considered -- to the full two particle
disorder vertex for the
conductivity in an otherwise non-interacting system. For the case of
 interacting Fermi systems, however, the presence of
the interactions is sufficient to destroy the aforementioned
 antisymmetry
and thus also the terms already included in the vertex of a single
 site two
particle CPA as discussed here will give a finite
contribution to the conductivity in real systems.

\subsection{Split band limit}

As already indicated, in contrast to many other theories of disorder,
 the CPA interpolates
correctly to the limits of strong disorder and high concentrations.
In this situation each band splits into two components of strength $c$
and $1-c$, respectively, which represent largely separate $A-$ and
 $B$-type
excitations. An intuitive consideration of the underlying physics
in this limit suggests, that the correct description of an absorption
 process should
predict that the overlap integral for transitions between sites
 pertaining to
different alloy components will gradually decrease and thus in
reverse, that transitions between
sites of equal type will be more and more favored. In the following,
 we prove that the CPA of weighted two particle
functions predicts this behavior correctly which makes it useful for a
better quantitative understanding of absorptive and dispersive
 processes
in strongly disordered alloys. The corresponding single particle
 theory must fail in this respect,
since it will weight the occurring transitions only with the
products of concentrations of  sites involved in these transitions.

To illustrate this we assume that our material components $A$
and $B$ have corresponding single site energies $
\varepsilon_a^A $, $ \varepsilon_b^A $ and $ \varepsilon_a^B $, $
\varepsilon_b^B $ for the two bands respectively and
the carriers have become totally localized, i.e. their effective mass
goes to infinity, or vice versa the bandwidths involved go to zero.
 The potentials $V_a$ and $V_b$ are then defined as
\begin{equation}
V_a = \varepsilon^A_a - \varepsilon^B_a, \hspace{20mm} V_b=
 \varepsilon^A_b - \varepsilon^B_b \label{sb1}
\end{equation}
and the single particle site diagonal Green's functions go over to
\begin{equation}
F_{\lambda}(z) = \frac c {z-\varepsilon_{\lambda}^A} + \frac {1-c}
 {z-\varepsilon_{\lambda}^B} \equiv \frac 1
 {z - \varepsilon_\lambda^B - \Sigma_{\lambda}} \label{sb2}
\end{equation}
where $\lambda$ labels the corresponding band.
Thus, if a two particle theory is constructed from single
particle properties only, and coherent terms in the two
particle scattering are neglected, this leads to peaks in
the absorption spectrum as shown in table \ref{Table1}. The energies
at which the peaks are centered are shown in row 2 and
their relative weight for the uncorrelated average is shown in the
row 3.

In the following we show for the limit of strong disorder how, upon
introducing the vertex corrections in conjunction with the appropriate
weighting factors for the two respective components, the expected
transitions are filtered out correctly with their appropriate
transition strengths and the spurious crossed terms are
 suppressed as shown in row 4 of table \ref{Table1}. Defining
\begin{equation}
z-\varepsilon_{\lambda}^B \equiv x \label{sb3}
\end{equation}
$F$ and $\Sigma$ can be rewritten as
\begin{eqnarray}
F(x) & = & \frac 1 {x- \Sigma} = \frac {x-(1-c)V} {x(x-V)}
 \label{sb4} \\
\Sigma(x)& = & \frac {cVx} {x-V(1-c)} \label{sb5}
\end{eqnarray}
and the vertex $\Lambda$ given by (\ref{ca17}) can be written
\begin{equation}
\Lambda = \frac {(1-c)c V_a V_b x_a x_b (x_a - V_a)(x_b-V_b)}
 { \left[ x_a - V_a(1-c) \right] \left[ x_b - V_b(1-c) \right]
 \left[ c x_a x_b + (1-c)(x_a-V_a)(x_b-V_b) \right]} \label{sb7}.
\end{equation}
The impurity weight $\xi \equiv (1 - c) \Lambda_{cv} /
 (V_c -\Sigma_c) (V_v - \Sigma_v)$ can be represented as
\begin{equation}
\xi  = \frac {cx_ax_b} {cx_ax_b + (1-c)(x_a - V_a)(x_b - V_b)}
 \label{sb8}
\end{equation}
and equally the host weight $\eta \equiv c \Lambda_{ab} / \Sigma_a
\Sigma_b$ as
\begin{equation}
\eta = \frac {(1-c)(x_a - V_a)(x_b - V_b)} {cx_ax_b +
(1-c)(x_a - V_a)(x_b - V_b)}. \label{sb9}
\end{equation}
It is now evident, that the impurity weighted
quantities are proportional to $c$ and the host weighted ones to
$1-c$ and not the other way around as their the appearance in terms
of their weighting factors might superficially have suggested.
The correction factor coming from the diagonal terms in the double
 weighted functions can be recast into
\begin{equation}
\gamma \equiv \frac {\Lambda F_a F_b} {V_a V_b} = \frac {(1-c) c} { c
x_a x_b + (1-c)(x_a-V_a)(x_b-V_b)} . \label{sb9.5}
\end{equation}
Furthermore, the unweighted but vertex corrected two particle
 propagator $K$ assumes the form
\begin{equation}
K= \frac { cx_ax_b + (1-c)(x_a - V_a)(x_b - V_b) } {x_ax_b
(x_a - V_a)(x_b - V_b) }. \label{sb10}
\end{equation}
After further algebra one can show, that in this limit the weighted
functions can be expressed as
\begin{eqnarray}
& & \langle G^{ii} G^{ii} \rangle = K \xi^2 + \gamma =
\langle G^i G^i \rangle = K \xi = \frac c {(x_a-V_a)(x_b-V_b)}
\label{sb11} \\
& & \langle G^{hh} G^{hh} \rangle = K \eta^2 + \gamma = \langle G^h
G^h \rangle = K \eta = \frac {(1-c)} {x_ax_b} \label{sb12} \\
& & \langle G^{ih} G^{ih} \rangle = 0 \label{sb13_5}
\end{eqnarray}
which is exactly what is expected to happen physically in this
limit. The crossed terms in the transition are canceled out --
hence the crossed function in (\ref{sb13_5}) goes to zero --
 and the double weighted
functions become identical to the single weighted ones, since
now effectively only the totally site diagonal element $K_D \equiv
 \langle G^{ \nu \mu
}(l,l) G^{ \nu \mu}(l,l) \rangle$ still contributes to the
 transitions, which implies that
only two possibilities for weighting the two particle functions
remain, namely as $\langle G^i_a G^i_b \rangle$ and $\langle G^h_a
G^h_b \rangle$. As could be expected from a theory which properly
describes the strong disorder limit the transition strengths now
distribute with the concentrations $c$ and $1-c$ between the $A_b
\rightarrow A_a$ and $B_b \rightarrow B_a$ transitions, respectively,
 such as
shown in row 4 of table \ref{Table1}.
This feature may in reverse  be used to derive the total diagonal
element $K_D$ for all ranges of disorder.
By requiring  $K^{ih}_D = 0$ we find
\begin{equation}
K_D = \frac \gamma {\xi\eta} = \frac {F_a F_b} {V_a V_b} \left( \frac
{ \Sigma_a \Sigma_b } c + \frac {(V_a - \Sigma_a ) (V_b - \Sigma_b)}
{(1-c)} \right) \label{sb13}.
\end{equation}
The weighted versions of this element are obtained by just multiplying
the corresponding single weights from (\ref{cc11}) and
(\ref{cc12}) on to it. Moreover, $K_D$ is equivalent to the
$r_{1,2}=0$ component of the two particle function
calculated for the finite range susceptibility in the last
sub-section. In terms of the
notation introduced there it reads
\begin{equation}
K_D = \sum_k \frac {a_k} {\Sigma_a \Sigma_b -
\Lambda_{a,b} a_k} \label{sb14}
\end{equation}
which would have been harder to evaluate starting from that
 representation.
The total site diagonal element thus
decouples into the corresponding site diagonal single particle
functions with an appropriate correction term accounting for the
 coherent
processes.

\section{Numerical Results}

In this subsection, in order to exemplify the general results, we
discuss numerical results obtained for the optical absorption in a
non-degenerate binary semiconductor alloy for a given  model density
 of states. We are thus able to show how a CPA type of
polarization, including vertex corrections, decomposes into
contributions originating from single alloy components as the
 strength of
the disorder is increased thus eventually causing the joint density
 of states
 to split into several components (up to three different ones for the
double weighted case).

In all our calculations we have used a semi elliptic
density of states for a pair of 3-D conduction and valence bands as
introduced in Ref. \onlinecite{bi3} for the single particle CPA, i.e.
\begin{equation}
\begin{tabular}{ c c }
$ \rho_{\lambda} (E) = \displaystyle{ \frac 2 {\pi w_{\lambda}^2}}
 \sqrt{w_{\lambda}^2 - E^2} $ &
$ \mid E \mid \leq w_{\lambda} $ \\
\\
$ \rho_{\lambda} (E) = 0 $ & $ \mid E \mid \leq w_{\lambda} $
 \label{n1}
\end{tabular}
\end{equation}
where  $\lambda$ labels either the conduction $\lambda = a$ or valence
$\lambda = b$, and $w_{\lambda}$ is the half-width of the band
 considered.
This has the advantage that the self consistent CPA equation for the
self energy $\Sigma (E)$ is a third degree polynomial which can be
 solved analytically.

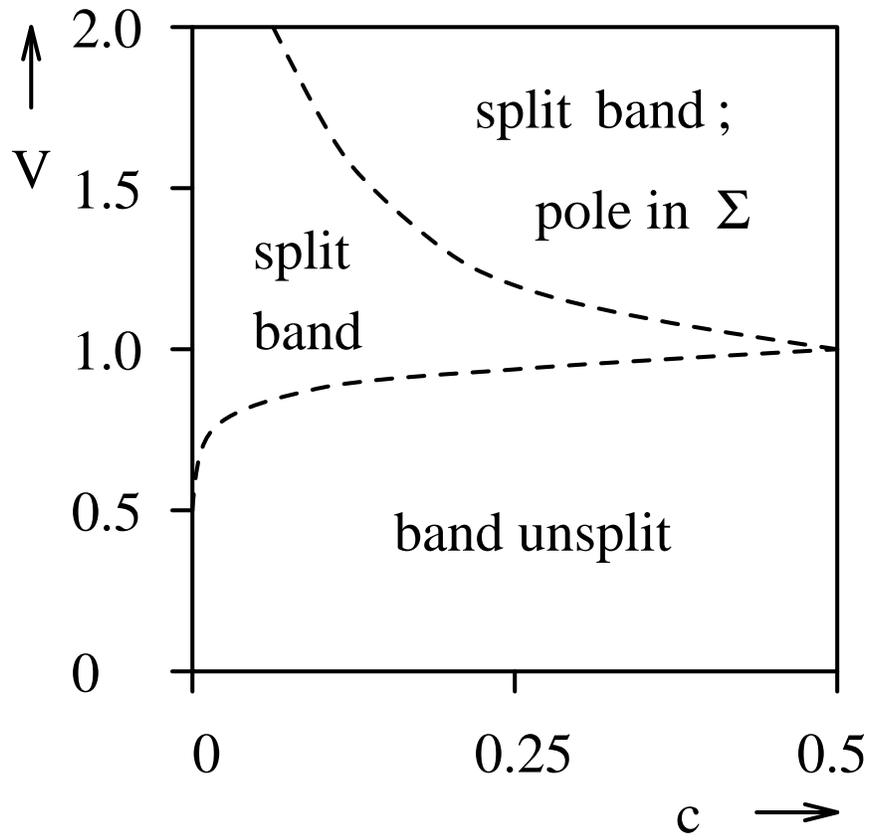
\begin{figure}[tbp]
\caption{Regimes for the CPA of the single particle density of
states (120) for $w_{\lambda}=1$ in a binary substitutional alloy
 depending on the impurity
concentration $c$ and the disorder strength $V$. In the lowest region
the band is unsplit. Above the first broken line the bands split into
two components and above the second broken line the CPA self energy
$\Sigma$ exhibits a pole between these components.}
\label{FIG1}
\end{figure}

To understand the effects that arise from genuine two particle
behavior as compared to those expected from the single particle CPA,
 we recapitulate some
of the features of the single particle theory first, mainly building
on the treatment presented in Ref. \onlinecite{bi3}. It is
 established there, that depending on the concentration and
disorder strength relative to the bandwidth, an impurity band is
eventually split off while in this split regime under some
circumstances the CPA self energy
exhibits a pole.
Fig. \ref{FIG1} shows a reproduction of the  ``phase'' diagram first
 presented
there, indicating  how the several
regions are separated. It can be seen that for a disorder
strength $ \mid V_{\lambda}/w_{\lambda} \mid > 1$ the bands always
 split into $A-$ and
$B$-components, whereas the splitting occurs earlier as  the
concentration $c$ is reduced, going
down to $\mid V_{\lambda}/w_{\lambda} \mid > 0.5$ as $c \rightarrow
 0$.

We have
calculated the linear polarizability of the medium by employing a Kubo
formula as introduced in (\ref{c1}). Furthermore,
 we continue assume that the optical matrix elements are
essentially constant and that such elements
are the same for both alloy components and we hence normalize
them to unity. The optical absorption is the negative imaginary part
 of the
retarded polarizability of the disordered medium
$-\text{Im}\Pi(\omega )
$, which can be formally written as
\begin{equation}
\Pi_r( \omega )  = - \lim_{\stackrel {\scriptstyle  \beta
\rightarrow \infty}{i \omega \rightarrow \omega + i \delta }}
\beta^{-1} \sum_{i \varepsilon} \int \frac {d^3 k}
{(2 \pi)^3} \langle G_a(k; i \varepsilon ) G_b(-k;i \omega - i
\varepsilon ) \rangle \label{n2}
\end{equation}
whereby the $k$-integration is understood to be carried out after
the configurational average has been performed, since before that both
resolvents would depend non-trivially on two momentum variables.

We consider our system at zero temperature and follow partly the
method used in Ref. \onlinecite{bi15} for our calculations.
At $T=0$, the polarization can be obtained as
the energy convolution around the conduction band branch cut of the
 $k$-summed vertex
corrected two particle function $\displaystyle{K(z_1,z_2) = \int
 \frac {d^3 k}
{(2 \pi)^3} \langle G_a(k;z_1 ) G_b(-k;z_2 ) \rangle}$, such that
\begin{equation}
\Pi_r( \omega ) = \oint_C K(z, \omega + i \delta - z) dz \label{n3}
\end{equation}
where we have taken over the following definitions from
Ref. \onlinecite{bi15}:
\begin{equation}
K(z_1,z_2) = \frac {R(z_1,z_2)} {1-\Lambda(z_1,z_2)R(z_1,z_2)}
 \label{n4}
\end{equation}
where $ \Lambda(z_1,z_2) $ is the usual CPA vertex from (\ref{ca17})
 and
\begin{equation}
R(z_1,z_2) = \int \frac  {d^3 k} {(2 \pi )^3} \bar G_a(k,z_1)
 \bar G_b(-k,z_2) \label{n5}
\end{equation}
is the average-decoupled two particle function.
Assuming that the conduction and valence band dispersion relations
exhibit a similar shape such that they scale proportionally
\begin{equation}
\frac { \varepsilon_a(k)} {w_a} = \mp \frac { \varepsilon_b(k)}
 {w_b} \label{n6}
\end{equation}
(\ref{n5}) can be shown to simplify to
\begin{equation}
R(z_1,z_2) = \frac {w_a F_a(z_1) \pm w_b F_b(z_2)} {w_a[z_2 -
\Sigma_b(z_2)] \pm w_b[z_1 - \Sigma_a(z_1)]} \label{n7}
\end{equation}
where $F_\lambda (z)$ are the site diagonal single particle functions
first introduced in connection with (\ref{b25}). As usual we assume
that the effective mass of an electron in the conduction band
is positive and that of a hole in the valence band is
negative. Accordingly we have chosen the upper choice of signs in
(\ref{n6}) and (\ref{n7}) for our calculations.

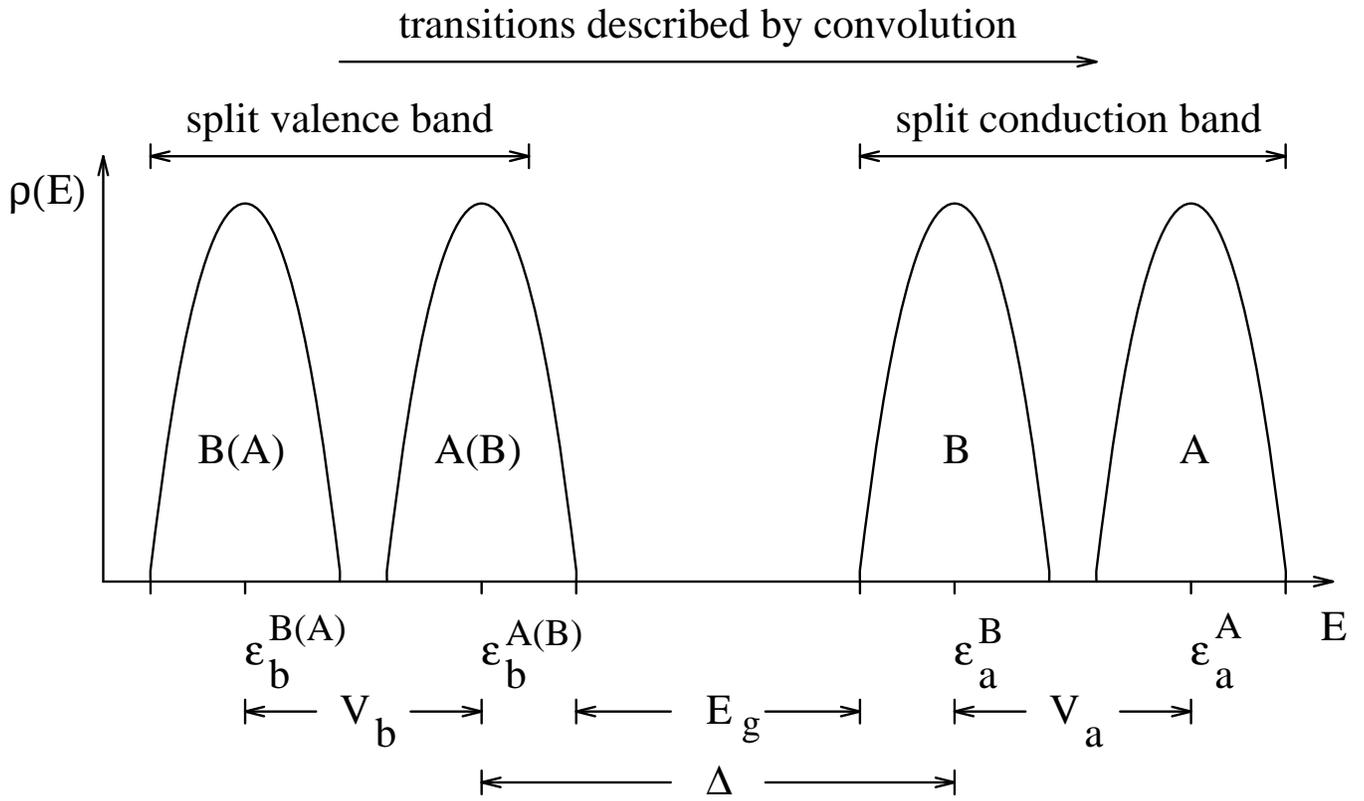
\begin{figure}[tbp]
\caption{Diagrammatic representation of the energy convolution in
(122) effectively using the term $R(z_1,z_2)$ from (126) only. If $R$
is taken at $z_1=z_2$ and $c$ is about $c=0.5$
the conduction and valence bands are split and
the imaginary part of $R$ decomposes into four components
centered about $\varepsilon^A_a$, $\varepsilon^B_a$ and
$\varepsilon^A_b$, $\varepsilon^B_b$ as shown qualitatively.
 The components are
separated by the band gap $E_g$ in the middle and the single particle
splittings in the conduction and valence parts at the upper and lower
end, respectively. The energetic
difference between the gravity centers are given by
$V_a=\varepsilon^A_a - \varepsilon^B_a$ and
$V_b=\varepsilon^A_b - \varepsilon^B_b$, as already introduced, as
 well as
$\Delta \equiv \varepsilon^B_a - \varepsilon^{A(B)}_b$. Bracketed
expressions denote anti-parallel disorder.}
\label{FIG2}
\end{figure}

To be able to analyze the obtained results with regard to the effect
of the the inclusion of vertex corrections we first consider
qualitatively the features that would be expected from the transition
process represented by the energy convolution in (\ref{n3}) in
an intermediate regime of disorder, if the configurational average
in the two particle function is decoupled and effectively only
single particle properties are employed. This would correspond to
replacing $K$ from (\ref{n4}) by $R$ from (\ref{n5})in (\ref{n3}).
 We assume
for now that the
concentration be about 0.5 and the bands have just split by a notable
amount. With the semi-elliptic bands used, the transition process can
be represented as shown in Fig. \ref{FIG2}.

The disorder strengths give approximately
the distance between the centers of the single bands.
The convolution of two
separated finite bands, occurring in $R$, would yield a set of finite
bands in the joint density of states (DOS) whose width is the sum of
 the widths of the contributing
components. Two cases are considered where the band offsets of the
$A-$ and $B$-components of the alloy are in equal or opposite
directions corresponding to parallel or anti-parallel disorder.

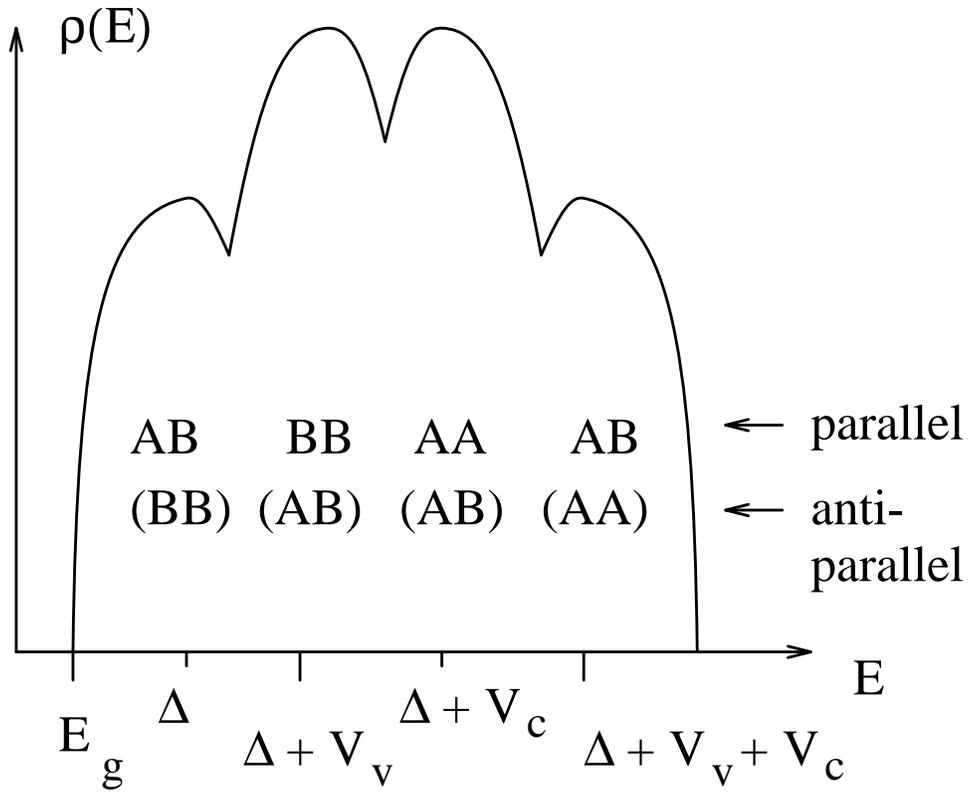
\begin{figure}[tbp]
\caption{
Qualitative result of the convolution (122) as it can be expected if
performed using the term $R(z_1,z_2)$ from (126) only.
 At about $c=0.5$ the
joint density of states would be expected to
distribute over regions in terms of their compositional origin as
 shown. The
inclusion of the vertex is expected to primarily suppress the mixed
$AB$-transitions and favor the $AA-$ and $BB$-transitions as the
disorder strength is increased. For parallel disorder this leads to a
suppression of the spectra at the flanks while for anti-parallel
disorder (in brackets) it causes a suppression of the central part.
 Both these
features are well represented in the calculated spectra of Figs. 4
 and 5.}
\label{FIG3}
\end{figure}

In the case of parallel disorder, this would amount to the $A_b
\rightarrow A_a$  and $B_b \rightarrow B_a$ -transitions lying in the
 center of the joint DOS, framed by the
contributions from the $B_b \rightarrow A_a$ and $A_b \rightarrow B_a$
-transitions upper and lower end respectively as shown in
Fig. \ref{FIG3}.
 At $c=0.5$ these
regions would have relative distribution of weighted states of
$1:2:1$ from lower end : center : higher end. In the case of
anti-parallel disorder the picture should be similar with the only
difference that the spectrum is turned inside-out with the $B_b
\rightarrow B_a$ and $A_b
\rightarrow A_b$ components on the top and the bottom end of the joint
DOS and the mixed transitions in the center, again with a
 distribution of
$1:2:1$.

The calculations in the previous section for the split band case
 strongly suggest that the
vertex corrections will increasingly suppress the cross transitions
as the disorder strength is increased, which is verified in our
numerical results. Indeed, our results show that this suppression is
 already
displayed quite strongly in an intermediate disorder range, i.e. in a
regime where the single bands just begin to split.

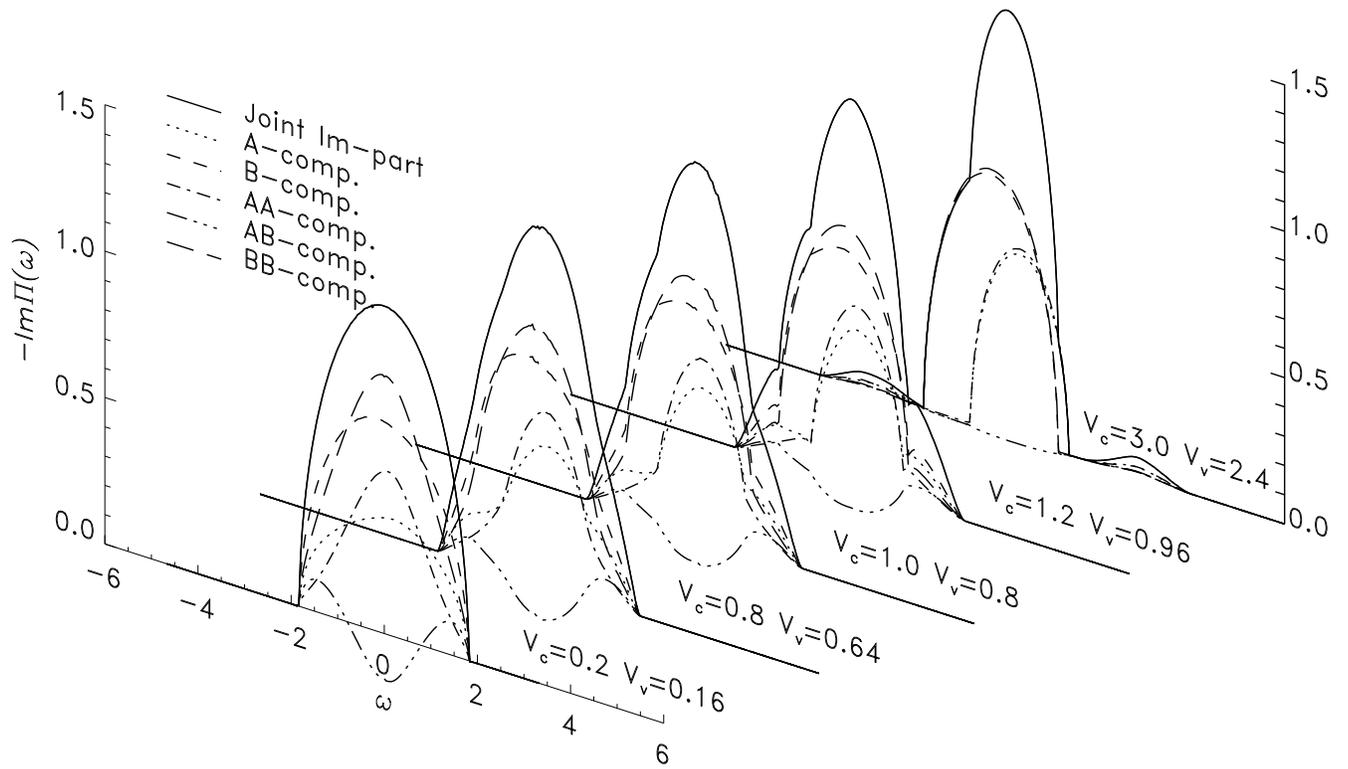
\begin{figure}[tbp]
\caption{Negative imaginary part of the polarizability of the
disordered medium (joint DOS)
for parallel disorder and its possible weightings into $A-$ and
$B-$ or
$AA-$, $AB-$ and $BB$-components. The concentration of $A$-atoms is
fixed to $c=0.35$ and the band half-widths were taken to be $w_a=1.0$
 and
$w_b=0.8$ for the conduction and valence bands, respectively. The
disorder was varied through a range of parameters, annotated on the
bottom right of the plots, with particular focus on the splitting
 point $ \mid
V_{\lambda}/w_{\lambda} \mid = 1$ of the single particle bands. At the
splitting point a pair of flank-components starts separating of
 sideways and gets
completely isolated as $V_{\lambda}/w_{\lambda} \rightarrow 3$ in the
last plot. These flanks are constituted to about $50\%$ by the
$AB$-component, whereas the central region almost entirely consists
 of $AA-$
and $BB$-components only.}
\label{FIG4}
\end{figure}

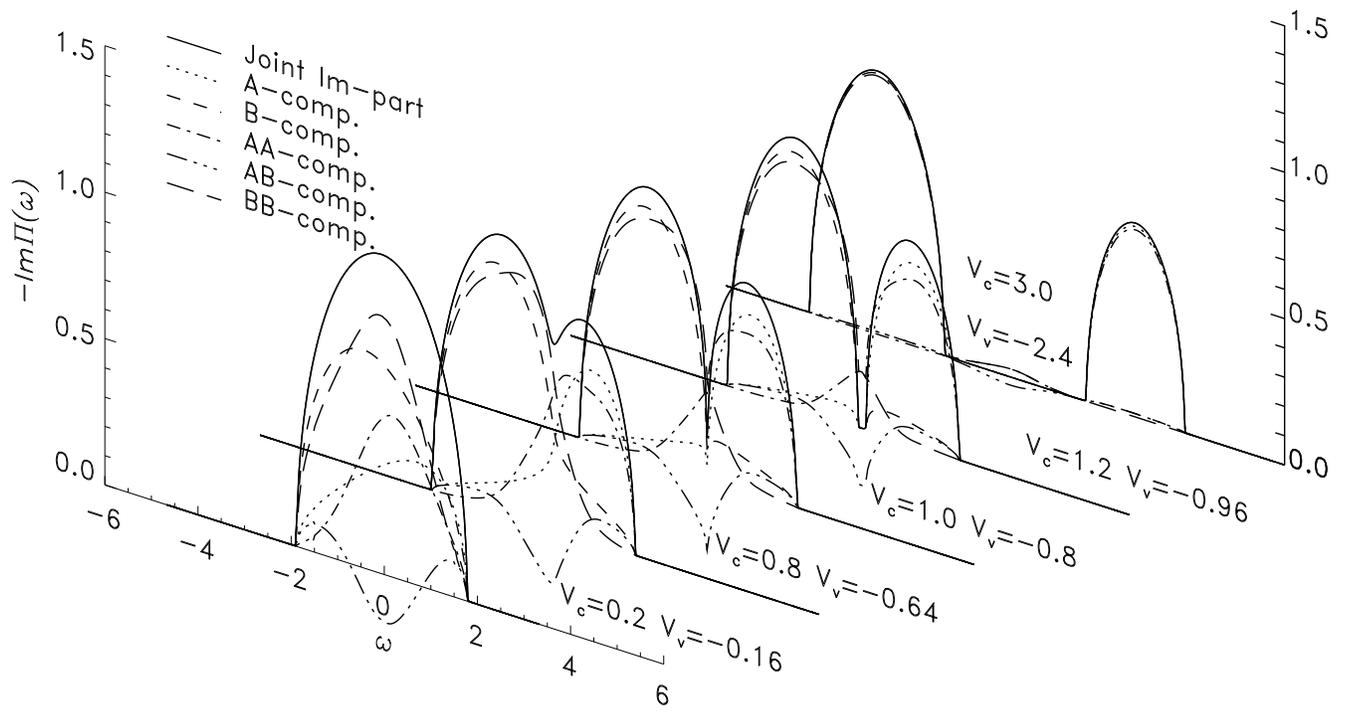
\begin{figure}[tbp]
\caption{Negative imaginary part of the polarizability for
anti-parallel disorder (reversed in the valence band) with otherwise
 equal model parameters to the parallel case shown in Fig. 4. The
 joint
part starts to exhibit a gap as soon as the conduction and valence
bands do. In the split regime the spectrum almost entirely builds up
from single $A-$ and $B$-components which in the interior of the two
bands again coincide largely with the $AA-$ and $BB$-components
indicating a strong breaking of the $k$-selection rule already in an
intermediate disorder regime. In the last plot $V_a=3.0$, $V_b=-2.4$
it can be seen that the double weighted functions are non-zero in the
gap region. This means that in a case where the optical cross sections
$\mu^{A/B}$ of the of $A-$ and $B$-atoms in (128) are non-equal, the
gap in the observed absorption spectrum would be less complete
 containing
either $AA-$ or $BB$-states primarily.}
\label{FIG5}
\end{figure}

The appended plots in Figs. \ref{FIG4} and \ref{FIG5} show cumulative
absorption spectra calculated from (\ref{n3}) as well as their single
 and
double weighted components for
parallel and anti-parallel disorder of various strengths,
covering both the joint and the split band regime. The conduction band
half-width is normalized to unity and the valence band half-width is
 taken to be $0.8$. The
concentration of impurities is fixed to $0.35$ in order to study
the
high-concentration behavior rather than dilute impurity effects.

In the transitional region when the disorder strengths start to exceed
the single particle half-bandwidths $ \mid
V_{ \lambda } / w_{ \lambda } \mid \geq 1$ and the conduction and
valence bands start to split we observe the following behavior: In
 the case of parallel
disorder shown in Fig. \ref{FIG4} the spectrum
starts to exhibit a discontinuity in its derivatives at the flanks
accounting for a
pair of mixed components splitting off sideways from the main
contribution. At the transition point the contributions of
the flanks relative to the central bulk part are approximately
 distributed in a ratio of $1:14:1$. In the case of anti-parallel
disorder shown in Fig. \ref{FIG5} the suppression of
the mixed transitions is even stronger so that as soon as the single
 bands
split the crossed
transitions cancel out entirely within the numerical accuracy
and the joint
density of states starts to exhibit a gap between two separate
contributions which mainly
consist of $A-$ and $B$-transitions, respectively.

In a regime of strong disorder, further beyond the splitting of the
single bands as shown in the last plots of Figs. \ref{FIG4} and
 \ref{FIG5}, one
finds that over large regions the single and uniformly double weighted
contributions coincide almost exactly, implying that
 the spectrum is built almost entirely from the total
diagonal element of the two particle function, which we had calculated
 at the end of section IV.
The total diagonal element can be obtained as an
independent $k$-sum over the two single particle resolvents involved.
This situation represents a breakdown of the $k$-selection rule
which holds in pure media.

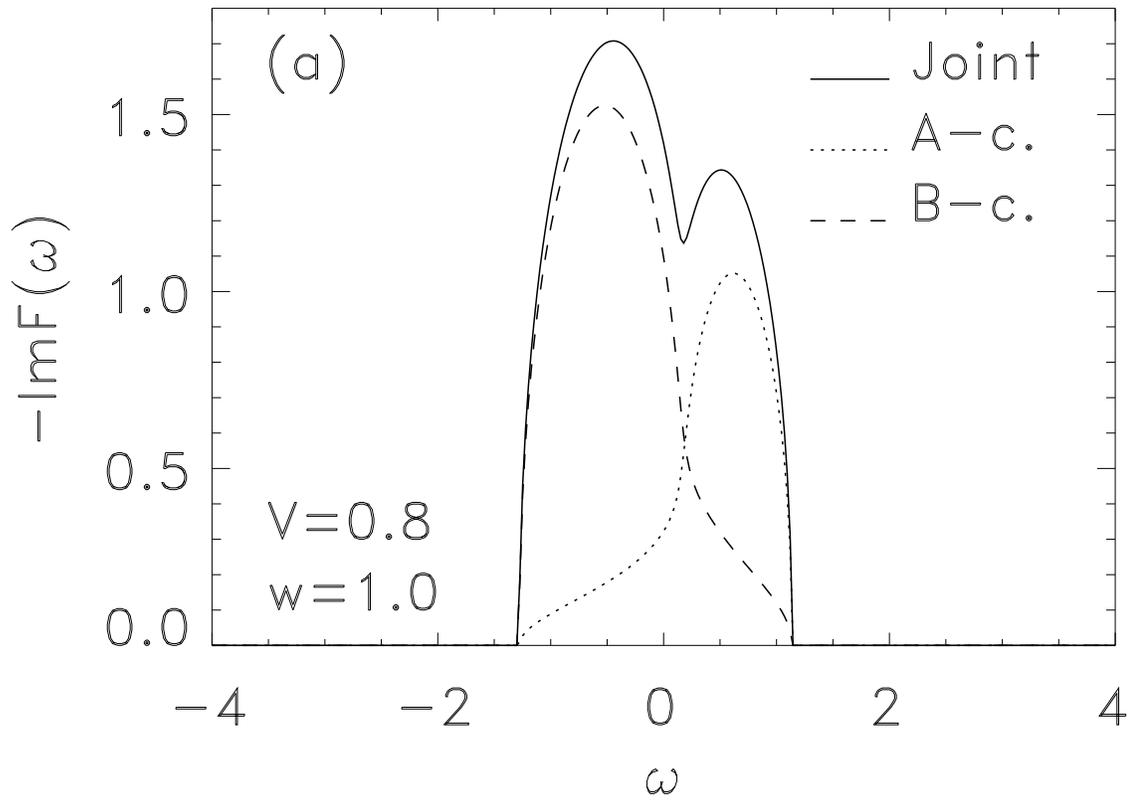
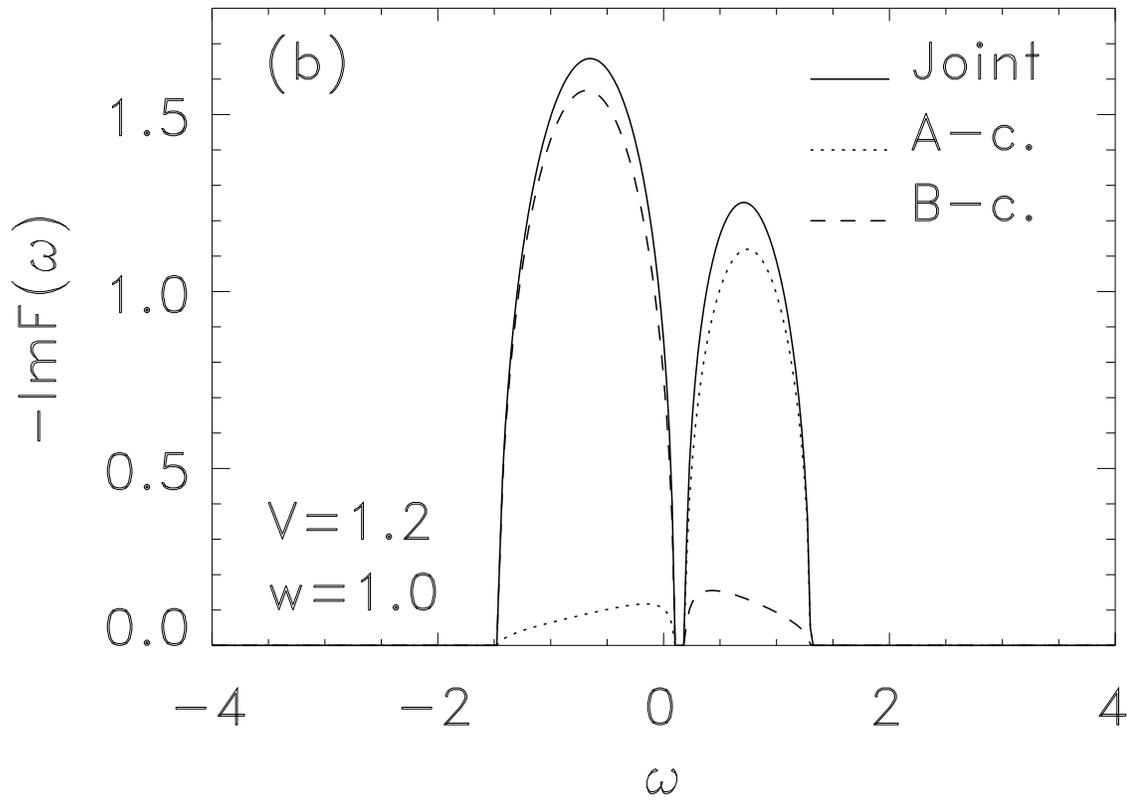
\begin{figure}[tbp]
\caption{Imaginary parts of the site diagonal single particle function
$F$ taken for the same concentration $c=0.35$, a band half-width
 of $w=1.0$
and disorder strengths of (a): $V = 0.8$ and (b): $V = 1.2$. By
 comparing
the splitting behavior of this function with the one of the
corresponding two particle function for anti-parallel disorder, shown
in the third and fourth plots in Fig. 5,
 one finds that the behavior, particularly the one of the single
 weighted
 $A-$ and $B$-components, is strikingly similar.}
\label{FIG6}
\end{figure}

If the splitting of $\Pi$ into single weighted components $\Pi =
 \Pi^A
+ \Pi^B $ is considered and the result is compared in appearance
 with the splitting into components of the site diagonal single
 particle function \cite{bi3}
\begin{equation}
F_r( \omega ) = \int \frac {d^3 k} {(2 \pi)^3} \bar G(k, \omega +
i \delta)
\end{equation}
one finds that the single and the two particle behavior appear to be
strikingly
similar in the case of anti-parallel disorder,
as can be seen if the plots for the imaginary part of the single
particle function shown in Figs. \ref{FIG6} (a) and (b) are compared
 to
the ones for the corresponding parameters in the two particle case
of Fig. \ref{FIG5} (third and fourth plot from the front).

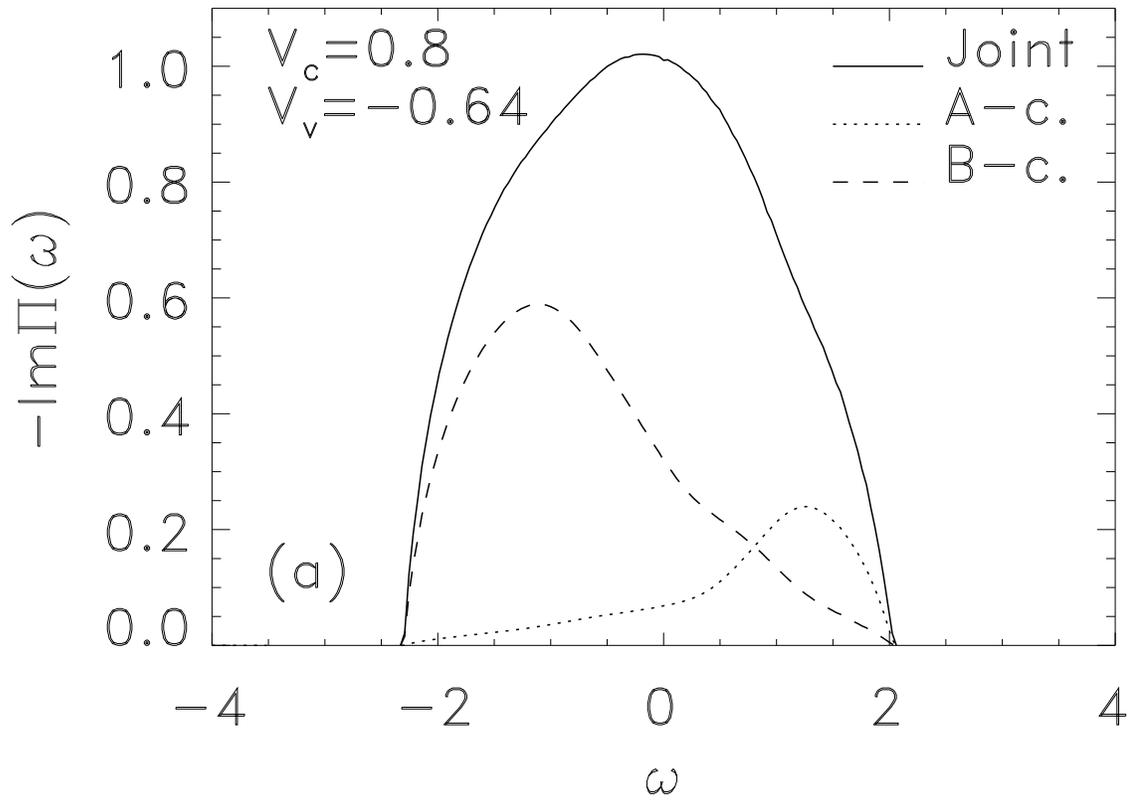
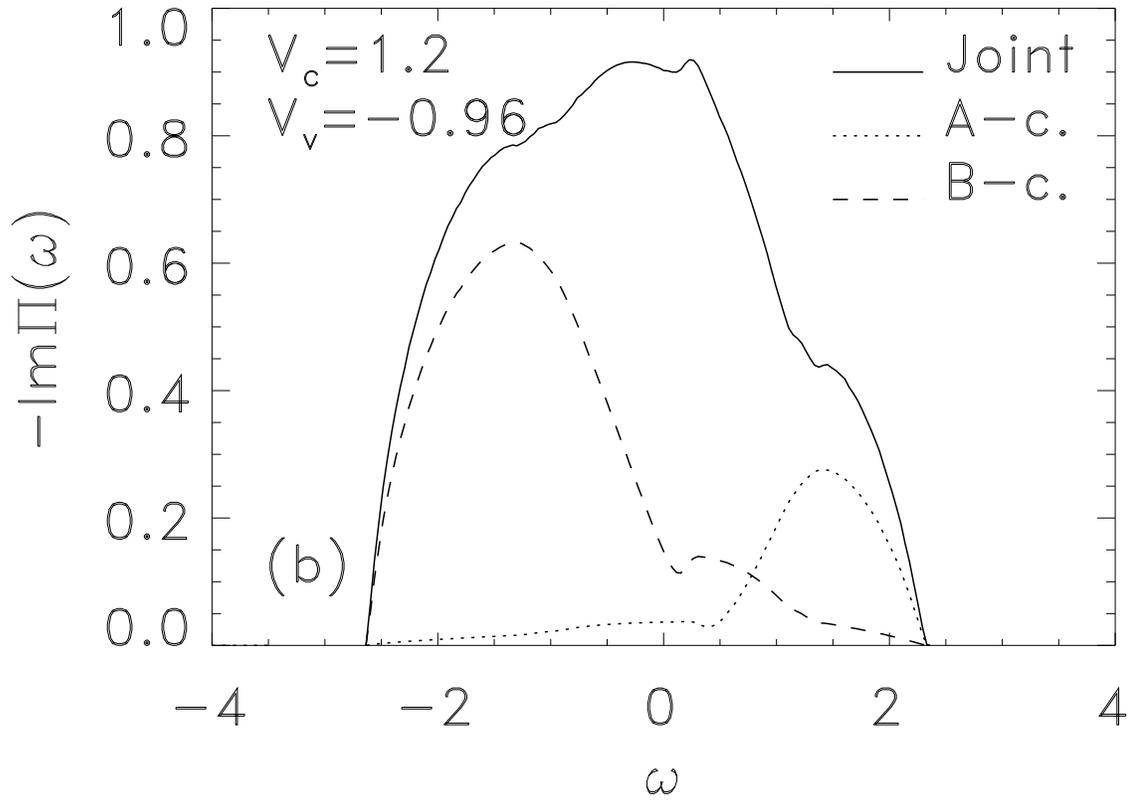
\begin{figure}[tbp]
\caption{Negative imaginary part of the polarizability without vertex
 corrections for
anti-parallel disorder. The parameters taken were (a): $c=0.35$,
 $V_a=0.8$,
$V_b=-0.64$ and (b): $c=0.35$, $V_a=1.2$, $V_b=1.0$. The
 characteristic
splitting of the joint DOS,  which is displayed in the properly
 corrected function of Fig. 5 as the disorder strengths exceed the
 value $ \mid V_{\lambda}/w_{\lambda} \mid = 1$, does not occur and
the shape of the joint DOS rather resembles the qualitative one of
Fig. 3. The curves
for the single weighted components in both (a) and (b), which were
obtained from correspondingly weighted uncoupled single particle
 functions, do not satisfy the sum rule $\Pi=\Pi^A + \Pi^B$.}
\label{FIG7}
\end{figure}

In a system without disorder this similarity is evident if
 there are no further local interactions in the problem,
since the non-interacting two particle motion decouples into a
center of mass and a relative co-ordinate and while the center of
mass motion can be set to zero, the relative one can be mapped onto a
single particle co-ordinate. Upon the addition of the disorder, this
decoupling fails to work
and it can only be regained by using an appropriate configurational
averaging procedure. However, if in the presence of disorder an
average is only performed on a single particle level, thus omitting
average induced two particle correlations the reduction obviously
fails to work, as is seen through comparing the plots of Fig.
 \ref{FIG7}
showing a
spectrum calculated without the vertex corrections with
 the ones for the corresponding parameter values of Fig. \ref{FIG5},
which properly include these corrections.
The results show, that in regimes of intermediate and strong
 disorder the influence of
the vertex corrections is very substantial.

In comparison to the single weighted (two fold) splitting the double
weighted (three fold) splitting exhibits a rather curious behavior.
 Even
though the components $-$Im$\Pi^{AA}$ and $-$Im$\Pi^{BB}$ lie
 underneath
their single weighted complements $-$Im$\Pi^{A}$ and $-$Im$\Pi^{B}$
in some
parts of the spectra, which one might expect to happen
globally on first thought, they either coincide with them or even
exceed them in other parts -- sometimes to such an extent that they
reach beyond the cumulative function. However, it has to be noted,
that these components, like the unweighted function, are always
uniformly positive in sign, and therefore exhibit the correct analytic
behavior that a function defined on this footing has to
satisfy. It is required that these components be positive
definite, because the net absorption in the medium
 must always be positive unless the
system is excited out of equilibrium, which we do not consider here.
In the preceding discussion we have
assumed for convenience that the optical matrix elements between
states of different components of the alloy are equal.
If different optical cross sections (matrix elements) $\mu^{A/B}$
 are
distinguished for $A-$ and $B$-atoms the weighted contributions to
yield the integrand of (\ref{n3}) would sum as
\begin{equation}
\tilde K= (\mu^A)^2 K^{AA} + 2\mu^A \mu^B K^{AB} + (\mu^B)^2 K^{BB}
 \label{n8}.
\end{equation}
This shows that it would be possible to observe one of the functions
$\Pi^{AA}$ or $\Pi^{BB}$ predominantly if either $\mu^A$ or $\mu^B$
 happens
to be much larger than the other. The mixed function $\Pi^{AB}$,
however,  will
never be a separately observable quantity in a general case, no
matter how the cross sections $\mu^A$ and $\mu^B$ scale
relatively and
therefore not so rigid criteria for its analyticity apply as for the
uniformly double weighted functions.

\section{Discussion and Conclusion}

In the previous sections we have obtained expressions for a wide class
of weighted two particle Green's functions.
The large choice for possible weightings is
substantially reduced as restrictions are made to functions which
would be useful in linear response theory. In both cases which
are discussed for this kind of application, only five
different weighted functions remain of which only two are genuinely
independent.

The structure of the weighting process is closely related to that
derived for the single particle theory with the main difference that
now the weights also depend significantly on the CPA vertex
 corrections.
The calculation for the split band limit, the domain in which the CPA
is
 superior to most of the other theories of disordered systems, gives
a direct insight into how the properly weighted CPA extracts the
correct limiting behavior from different possible physical processes.

Some care is needed in interpreting the precise physical meaning of
the weights, since they are obtained for the averaged functions,
 which describe the disordered medium as effectively
homogeneous. The concept of the propagation of a particle between
 sites
of different components is therefore lost in the effective
medium as a consequence of averaging and the initial exclusion of
 specific
propagation paths in the unaveraged function leads to the effective
weights. These weights
simply account for the average partition in probability for the
simultaneous propagation of the particles between partly or completely
specified site types at a given pair of energies $z_1$ and $z_2$.

Our numerical results show the general importance of the inclusion of
vertex corrections into a properly self consistent two particle
formalism. We managed to visualize that, as a consequence of the
 inclusion
of these average induced two particle correlations, the center of mass
and relative
motion of the two particle system effectively decouple to a large
 extent.

We believe that the general method developed here will find
 applications
in various situations where two particle motion is studied in a
disordered medium.
The effect of alloying on the electronic susceptibility and hence for
example on the
Rudermann-Kittel interaction has already been mentioned. In
particular we believe it can be extended for use in systems where the
two particles have a direct interaction, such as the Coulomb
interaction between carriers occurring in excitons in alloyed systems.
In such a case the static correlations between particles, created by
the disorder and accounted for by the vertex corrections, and the
dynamic correlations introduced through the carrier-carrier
interaction, create additional
static-dynamic correlations. Moreover it may
be possible that the underlying disorder of the system gives rise to
an induced disorder to the carrier-carrier interaction itself. Both of
these effects can be treated within the method developed here
and will be addressed in a forthcoming publication.

\section{Acknowledgements}

The authors would like to thank Mr. Carsten Heide at Oxford for
 reading the manuscript and making useful suggestions.

\begin{table}
\begin{center}
\begin{tabular}{ccccc}
Transition & $A_b \rightarrow A_a$ & $B_b \rightarrow A_a$ & $A_b
\rightarrow B_a$ & $B_b \rightarrow B_a$ \\
Energy of pole & $\varepsilon_a^A - \varepsilon_b^A$ &
$\varepsilon_a^A - \varepsilon_b^B$ & $\varepsilon_a^B -
\varepsilon_b^A$ & $\varepsilon_a^B - \varepsilon_b^B$ \\
 Weight without vertex & $c^2$ & $c(1-c)$ &
$c(1-c)$ & $(1-c)^2$ \\
Weight including vertex & $c$ & 0 & 0 & $1-c$
\end{tabular}
\end{center}
\caption{Energies of poles at which interband transitions may occur
versus center of gravity weights in an asymptotically strong disorder
limit before and after the inclusion of vertex corrections.}
\label{Table1}
\end{table}

\end{document}